\documentstyle[agupp,psfig]{article} 
\lefthead{HELMSTETTER and SORNETTE}
\righthead{Predictability of Triggered Seismicity}	
        \authoraddr{Agn\`es Helmstetter, Laboratoire de G{\'e}ophysique
Interne et Tectonophysique,
         Observatoire de Grenoble, Universit{\'e} Joseph Fourier, BP 53X, 38041
         Grenoble Cedex, France. (e-mail: ahelmste@obs.ujf-grenoble.fr)}
\authoraddr{Didier Sornette,
Department of Earth and Space Sciences and Institute of
Geophysics and Planetary Physics, University of California, Los Angeles,
California and Laboratoire de Physique de la Mati\`{e}re Condens\'{e}e,
CNRS UMR 6622 and Universit\'{e} de Nice-Sophia Antipolis, Parc Valrose,
06108 Nice, France (e-mail: sornette@moho.ess.ucla.edu)}
\input{update.tex}
\begin{document}
\title{Predictability in the ETAS Model of Interacting Triggered
Seismicity}
\author{Agn\`es Helmstetter}
\affil{Laboratoire de G{\'e}ophysique Interne et Tectonophysique,
        Observatoire de Grenoble, Universit\'e Joseph Fourier, France}
\author{Didier Sornette}
\affil{Department of Earth and Space Sciences and Institute of
Geophysics and Planetary Physics, University of California, Los Angeles,
       California 90095-1567 and
     Laboratoire de Physique de la Mati\`{e}re Condens\'{e}e, CNRS UMR 6622
Universit\'{e} de Nice-Sophia Antipolis, Parc Valrose, 06108 Nice, France
}

\newcommand{\be}{\begin{equation}}
\newcommand{\ee}{\end{equation}}
\newcommand{\ba}{\begin{eqnarray}}
\newcommand{\ea}{\end{eqnarray}}
\newenvironment{technical}{\begin{quotation}\small}{\   end{quotation}}

\begin{abstract}

As part of an effort to develop a systematic methodology for earthquake
forecasting, we use a simple model of seismicity based on interacting events
which may trigger a cascade of earthquakes, known as the Epidemic-Type
Aftershock Sequence model (ETAS). The ETAS model is constructed on
a bare (unrenormalized) Omori law, the Gutenberg-Richter law and the
idea that large events trigger more numerous aftershocks. For simplicity,
we do not use the information on the spatial location of earthquakes
and work only in the time domain. We demonstrate the essential role 
played by the cascade of triggered seismicity in
controlling the rate of aftershock decay as well as the overall level
of seismicity in the presence of a constant external seismicity source.
We offer an analytical approach to account for the yet unobserved
triggered seismicity adapted to the problem of forecasting future
seismic rates at varying horizons from the present.
Tests  presented on synthetic catalogs validate strongly
the importance of taking into account all the cascades of
still unobserved triggered events in order to predict correctly the 
future level of seismicity beyond a few minutes.
We find a strong predictability if one accepts to predict only
a small fraction of the large-magnitude targets. Specifically, 
we find a prediction gain (defined as the ratio of the fraction
of predicted  events over the fraction of time in alarms) equal to 21
for a fraction of alarm of 1\%, a target magnitude $M\geq6$, an 
update time of 0.5 days between two predictions, and for realistic parameters
of the ETAS model.
However, the probability gains degrade fast when one attempts
to predict a larger fraction of the targets. This is because
a significant fraction of events remain uncorrelated from past seismicity.
This delineates the fundamental limits underlying forecasting skills, 
stemming from an intrinsic stochastic component in these interacting 
triggered seismicity models. Quantitatively, the fundamental limits 
of predictability found here are only lower bounds of the true values 
corresponding to the full information on the spatial location of earthquakes.

\end{abstract}

\begin{article}

\section{Introduction}

There are several well documented facts in seismicity:
(1) spatial clustering of earthquakes at many scales,
[e.g. {\it Kagan and Knopoff}, 1980]
(2) the Gutenberg-Richter (GR) [{\it Gutenberg and Richter}, 1944]
distribution of earthquake magnitudes
and (3) clustering in time following large earthquakes,
quantified by Omori's $\approx 1/t^p$ law for aftershocks
(with $p \approx 1$) [{\it Omori}, 1894].
However, there are some deviations from these empirical laws, and
a significant variability in the parameters of these laws.
The $b$-value of the GR law, the $p$ Omori exponent and the
aftershock productivity are spatially and temporally variable
[{\it Utsu et al.}, 1995; {\it Guo and Ogata}, 1997].
Some alternative laws have also been proposed, such as a gamma law
for the magnitude distribution [{\it Kagan}, 1999] or
the stretched exponential for the temporal decay of the rate of
aftershocks [{\it Kisslinger}, 1993].

However, these ``laws'' are only the beginning of a full model of seismic
activity and earthquake triggering.
In principle, if one could obtain a faithful representation (model)
of the spatio-temporal organization of seismicity, one could use
this model to develop algorithms for forecasting earthquakes.
The ultimate quality of these forecasts would be limited by the quality
of the model, the amount of data that can be used in the forecast and its
reliability and precision, and the stochastic component of seismic activity.
Here, we analyze a simple model of seismicity known as the Epidemic Type 
Aftershock Sequence model (ETAS) [{\it Ogata}, 1988, 1989] and we use it
to test the fundamental limits of predictability of this class of models.
We restrict our analysis to the time domain, that is, we neglect the
information provided by the spatial location of earthquakes which could
be used to constrain the correlation between events and would be expected
to improve forecasting skills. Our results should thus give lower
bounds of the achievable predictive skills. This exercise is rather
constrained but turns out to provide meaningful and useful insights.
Because our goal is to estimate the intrinsic limits of
predictability in the ETAS model, independently of the additional
errors coming from the uncertainty in the estimation of the ETAS
parameters in real data, we consider only synthetic catalogs generated
with the ETAS model. Thus, the only source of error of the prediction
algorithms result from the stochasticity of the model.

Before presenting the model and developing the tools necessary for
the prediction of future seismicity, we briefly summarize in the
next section the available methods of earthquake forecasting based on past 
seismicity. In section 3, we present the model
of interacting triggered seismicity used in our analysis.
Section 4 develops the formal solution
of the problem of forecasting future seismicity rates based
on the knowledge of past seismicity quantified by a catalog of
times of occurrences and magnitudes of earthquakes. Section 5 gives
the results of an intensive series of tests, which quantify in several 
alternative ways the quality of forecasts (regression of predicted 
versus realized seismicity rate, error diagrams, probability gains, 
information-based binomial scores). Comparisons with the Poisson 
null-hypothesis give a very significant success rate. However, only
a small fraction of the large-magnitude targets can be shown to be successfully
predicted while the probability gain deteriorates rapidly when one attempts
to predict a larger fraction of the targets. We provide a detailed
discussion of these results and of the influence of the model
parameters. Section \ref{conclu} concludes.

\section{A rapid tour of methods of earthquake forecasts based on
past seismicity}

All the algorithms that have been developed for the prediction of future
large earthquakes based on past seismicity rely on their
characterization either as witnesses or as actors. In other
words, these algorithms assume that past seismicity
is related in some way to the approach of a large scale rupture.

\subsection{Pattern recognition (M8)}

In their pioneering  work, {\it Keilis-Borok and  Malinovskaya} [1964] 
codified an observation of general increase in seismic activity
depicted with the only measure, ``pattern Sigma'', which is characterizing 
the trailing total sum of the source areas of the medium size 
earthquakes. Similar strict codifications of other seismicity patterns,  
such as a decrease of $b$-value, an increase in the rate of activity, an 
anomalous number of aftershocks, etc, have been proposed later and define 
the M8 algorithm of earthquake forecast (see [{\it Keilis-Borok and Kossobokov}, 
1990a; {\it Keilis-Borok and Soloviev}, 2002] for useful reviews). 
 In these algorithms, an alarm is defined when several
precursory patterns are above a threshold calibrated in a training period.
Predictions are updated periodically as new data become available.
Most of the patterns used by this class of algorithms are reproduced by the
model of triggered seismicity known as the ETAS (Epidemic Type
Aftershock Sequence) model [see {\it Sornette and Sornette}, 1999;
{\it Helmstetter and Sornette}, 2002; {\it Helmstetter et al.}, 2003].
The prediction gain $G$ of the M8 algorithm, defined as the ratio between the
fraction of predicted events over the fraction of time occupied by alarms,
is usually in the range 3 to 10 (recall that a random predictor would give
$G=1$ by definition).
A preliminary forward test of the algorithm for the time period July 1991
to June 1995 performed no better than the null hypothesis using a reshuffling
of the alarm windows [{\it Kossobokov et al.,} 1997].
Later tests indicated however a confidence level of 92\%
for the prediction of $M7.5+$ earthquakes by the algorithm M8-MSc
for real-time intermediate-term predictions in the Circum Pacific seismic belt,
1992-1997, and above 99\% for the prediction of $M\geq8$ earthquakes
[{\it Kossobokov et al.}, 1999]. We use the term confidence level
as $1$ minus the probability of observing a predictability
at least as good as what was actually observed, under the null hypothesis
that everything is due to chance alone.
As of July 2002, the scores (counted from the formal start of
the global test initiated by this team since July 1991) are as follows:
For $M8.0+$,  8 events occurred, 7 predicted by M8, 5 predicted by M8-MSc;
for $M7.5+$,  25 events occurred, 15 predicted by M8 and 7 predicted by M8-MSc.

\subsection{Short term forecast of aftershocks}

{\it Reasenberg and Jones} [1989] and {\it Wiemer} [2000] have developed
algorithms to predict the rate of aftershocks following major earthquakes.
The rate of aftershocks of magnitude $m$ following an earthquake of
magnitude $M$ is estimated by the expression
\be
N_M(m)={ k~10^{b(M-m)} \over (t+c)^p}~,
\label{sfde}
\ee
where $b \approx 1$ is the $b$-value of the Gutenberg-Richter (GR) distribution.
This approach neglects the contribution of seismicity prior to the
mainshock, and does not take into account the specific times, locations
and magnitudes of the aftershocks that have already occurred.
In addition, this model (\ref{sfde}) assumes arbitrarily that the rate
of aftershocks  increases with the magnitude $M$ of the mainshock as
$\sim 10^{b M}$, which may not be correct.
A careful measure of this scaling for the southern California seismicity gives
a different scaling $\sim 10^{\alpha M}$ with $\alpha=0.8$
[{\it Helmstetter}, 2003]. Moreover, an analytical study of the ETAS model
[{\it Sornette and Helmstetter}, 2002] shows that the  case $\alpha \geq b$
leads to an explosive seismicity rate,  which is unrealistic to describe
the seismic activity.

\subsection{Branching models}

Simulations using branching models as a tool for predicting
earthquake occurrence over large time
horizons were proposed in {\it Kagan} [1973], and first implemented in
{\it Kagan and Knopoff} [1977]. In a recent work, {\it Kagan and
Jackson} [2000] use a variation of the ETAS model to estimate
the rate of seismicity in the future but they neglect
the seismicity that will be triggered between the present
time and the horizon and which may dominate the future activity.
Therefore, these predictions are only valid at very short times, when
very few earthquakes have occurred between the present and the horizon.

To solve this problem and to extend the predictions further in time,
{\it Kagan and Jackson} [2000] propose to use Monte-Carlo simulations
to generate many possible scenarios of the future seismic activity.
However, they do not use this method in their forecasting procedure.
These Monte-Carlo simulations  will be implemented in our tests,
as we describe below. This method has already been tested
by {\it Vere-Jones} [1998] to predict a synthetic catalog generated
using the ETAS model.
Using a measure of the quality of seismicity forecasts in terms of
mean information gain per unit time, he obtains scores usually
worse than the Poisson method.
We use below the same class of model and implement a procedure
taking into account the cascade of triggering.
We find, in contrast with the claim of  {\it Vere-Jones} [1998],
a very strong probability gain. We explain in section 5.6 the origin
of this discrepancy, which can be attributed to the use of different measures
for the quality of predictions. 

In [{\it Helmstetter et al.}, 2003],
the forecasting skills of algorithms based on three functions of
the current and past seismicity (above a magnitude threshold)
measured in a sliding window of 100 events
have been compared. The functions are
(i) the maximum magnitude $M_{max}$ of the 100 events
in that window, (ii) the  Gutenberg-Richter $b$-value measured
on these 100 events by the standard Hill maximum likelihood estimator and
(iii) the seismicity rate $\lambda$ defined as the inverse of the duration
of the window. For each function, an alarm was declared for the target of an earthquake
of magnitude larger than $6$
when the function is either larger
(for $M_{max}$ and $\lambda$) or smaller (for $b$) than a threshold.
These functions $M_{max}$, $b$ and $\lambda$ are similar and in some
cases identical
to precursors and predictors that have been studied by other authors.
{\it Helmstetter et al.} [2003] found that these three predictors
are considerably better than those obtained for a random prediction, with the
prediction based on the seismicity rate $\lambda$ being by far the best. This is
a logical consequence of the model of interacting triggered seismicity
used in {\it Helmstetter et al.} [2003] and also in the present work,
in which any relevant physical observable is a function of the seismicity rate.
At least in the class of interacting triggered seismicity, the largest
possible amount of information is recovered by targeting the seismicity rate.
All other targets are derived from it as linear or non-linear
transformations of it. Our present study refines and extends the
preliminary tests of {\it Helmstetter et al.} [2003] by
using a full model of seismicity rather than the coarse-grained
measure $\lambda$. We note also that the forecasting methods of {\it Rundle et al.}
[2001; 2002] are based on a calculation of the coarse-grained seismicity above
  a small magnitude threshold, which is then projected into the future.

\section{The model of triggered seismicity}

The parametric form that defines the ETAS model used in this paper
was formulated by {\it Ogata}  [1985, 1987, 1988].
See [{\it Ogata}, 1999] and [{\it Helmstetter and Sornette}, 2002a] for reviews
of its origins, a description of the different versions of the model and of its
applications to model or predict seismic activity.
It is important to stress that the ETAS model
is not only a model of aftershock sequences as the acronym ETAS
(Epidemic-Type Aftershock Sequence) would make one to believe but is
fundamentally a model of triggered interacting seismicity.

In addition to the strict definition of the ETAS model used by
{\it Ogata} [1985, 1987, 1988, 1989, 1999], there were and still are a
variety of alternative parametric forms of the extended
``mutually exciting point processes'' with marks (that is, magnitudes)
introduced by {\it Hawkes} [1971, 1972], which have
been applied to earthquakes, including {\it Kagan and Knopoff} [1987];
{\it Kagan} [1991] and {\it Lomnitz} [1974].
{\it Kagan and Knopoff} [1987] differs from {\it Ogata} [1985, 1987, 1988]
in replacing the role played by the parameter $c$ in the modified Omori law
(\ref{sfde}) by an abrupt cut-off which models the duration of the mainshock.
They think that a non-zero value of $c$ is merely the artifact of the
missing events immediately after the mainshock
In contrast, based on the observation of the records of seismic waves,
{\it Utsu} [1970, 1992] considers that the parameter $c$ is not merely
due to such artifact but also possesses some physical meaning.
The analysis of {\it Helmstetter and Sornette} [2002a] shows that
the choice of a non-zero $c$ value [{\it Ogata}, 1988] or an abrupt cut-off
[{\it Kagan and Knopoff}, 1987] does not lead to any detectable differences in
simulated catalogs at time scales beyond $c$ (which is usually very small).
Thus, from the point of view of the collective behavior of the model,
both formulations lead to essentially indistinguishable catalogs and
statistical properties. {\it Lomnitz} [1974]'s model (``Klondike
model'') was also directly inspired by {\it Hawkes} [1971]
and is similar to the ETAS model, but assumes different parametric
forms: in particular, the number of triggered events is taken
proportional to the magnitude, not the exponential of the magnitude.
{\it Kagan and Jackson} [2000] also use a formulation of the same class
but with a more complex specification of the time, space and
magnitude dependence of the triggering process and propagator.

\subsection{Definition of the ETAS model}

The ETAS model of triggered seismicity is defined as follows
[{\it Ogata}, 1985; 1987; 1988; 1989; 1992; 1999].
We assume that a given event (the ``mother'') of magnitude
$m_i \geq m_0$ occurring at time $t_i$  and position $\vec r_i$
gives birth to other events (``daughters'') in the time interval
between $t$ and $t+dt$ and at point $\vec r \pm \vec dr$  at the rate
\be
\phi_{m_i}(t-t_i, \vec r-\vec r_i)
= \rho(m_i) ~\Phi(t-t_i)~\Phi(\vec r-\vec r_i)~.
\label{eq1}
\ee

$\Phi(t)$ is the direct Omori law normalized to $1$
\be
\Phi(t) = {\theta~c^{\theta} \over (t+c)^{1+\theta}}~H(t)~,
\label{Phi}
\ee
where $\theta>0$, $H(t)$ is the Heaviside function, and $c$ is a regularizing
time scale that ensures that the seismicity rate remains finite close to the
mainshock.

$\Phi(\vec r-\vec r_i)$ is a normalized spatial ``jump'' distribution
from the mother to each of her daughter, which quantifies the probability
for a daughter to be triggered at a distance $|\vec r-\vec r_i|$ from
the mother, taking into account the spatial dependence of the
stress induced by an earthquake.

$\rho(m)$ gives the total number of aftershocks triggered directly
by an event of magnitude $m$
\be
\rho(m)=k~10^{\alpha (m-m_0)}~,
\label{rho}
\ee
where  $m_0$ is a lower bound magnitude below which no daughter is triggered.
The adjective ``direct'' refers to the events of the first generation triggered
in first-order lineage from the mother event. The formulation of
(\ref{Phi}) and (\ref{rho}) is originally due to {\it Utsu} [1970].

The model is complemented by the Gutenberg-Richter (GR) law which states that
each earthquake has a magnitude chosen according to the
density distribution
\be
P(m) = b~\ln(10)~10^{-b(m-m_0)}~.
\label{GRlaw}
\ee
$P(m)$ is normalized: $\int_{m_0}^{\infty}  ~P(m)~dm =1$.
When magnitudes are translated into energies, the GR law becomes the
(power law) Pareto law.
See [{\it Ogata}, 1989; 1992; 1999; {\it Guo and Ogata}, 1997; {\it Helmstetter
and Sornette}, 2002a] for a discussion of the values of the ETAS
parameters $c$, $\alpha$, $\theta$, $K$ and $b$ in real seismicity.
Note however that we have reasons to think that previous inversions
of the ETAS parameters in real data are unreliable due to lack of
concern for the unobserved seismicity below the completeness threshold.
We are presently working on methods to address this question. 

In this first investigation, we limit ourselves to the time domain,
studying time series of past seismicity summed over an overall spatial region,
without taking into account information on earthquake locations.
This amounts to integrating the local Omori law (\ref{eq1}) over the
whole space.
In the following, we will thus use the integrated form of the  local Omori
law (\ref{eq1}) given by
\be
\phi_{m_i}(t-t_i)= \int \limits_r \phi_{m_i}(t-t_i,\vec r-\vec r_i)~dr
= \rho(m_i)~\Phi(t-t_i)~.
\label{eq2}
\ee
The complete model with both spatial and temporal dependence (\ref{eq1})
has been studied in [{\it Helmstetter and Sornette}, 2002b] to derive the joint
probability distribution of the times and locations of aftershocks including
the whole cascade of secondary and later-generations 
aftershocks. When integrating the rate of
   aftershocks calculated for the spatio-temporal ETAS model over the whole
space, we recover the results used in this paper for the temporal
ETAS model.

We stress that not taking into account the spatial positions of the
earthquakes is not saying that earthquakes occur at the same position.
The synthetic catalogs we generate in space and time are similar
to real catalogs and our procedure just neglects the information on space.
Clearly, this is not what a full prediction method should do and it is clear
that not using the information on the location of earthquakes
will underestimate (sometimes grossly) the
predictive skills that could be achieved with a full spatio-temporal
treatment.
However, the problem is sufficiently complex that we find it useful to
go through this first step and develop the relevant concepts and first tests
using only information on seismic time sequences.

\subsection{Definition of the average branching ratio $n$}

The key parameter of model (\ref{eq1}) is the average number
       (or ``branching ratio'') $n$ of daughter-earthquakes
created per mother-event. This average is performed over time and over
all possible mother magnitudes.
This average branching ratio $n$ is a finite value for $\theta>0$
and for $\alpha<b$, equal to
\be
n \equiv \int \limits_0^{\infty} dt
\int \limits_{m_0}^{\infty}~P(m)~\rho(m)~\Phi(t)~dm =
{ k b \over b-\alpha}.
\label{nvaqlue}
\ee
The normal regime corresponds to the subcritical case $n<1$ for which
the seismicity rate decays after a mainshock to a constant level (in
the case of a steady-state source) with fluctuations in the seismic
rate.

Since $n$ is defined as the average  over all mainshock magnitudes
of the mean number of events triggered by a mainshock,
the branching ratio does not give the number of daughters
of a given earthquake, because this number also depends on the
specific value of its magnitude as shown by (\ref{rho}).
As an example, take $\alpha=0.8$, $b=1$, $m_0=0$ and $n=1$. Then, a
mainshock of magnitude $M=7$ will have on average 80000 direct aftershocks,
compared to only 2000 direct aftershocks for an earthquake of magnitude $M=5$
and less than  0.2 aftershocks for an earthquake of magnitude $M=0$.

The branching ratio defined by (\ref{nvaqlue}) is the key parameter of
the ETAS model, which controls the different regimes of seismic activity.
There are two observable interpretations for this parameter
[{\it Helmstetter and Sornette}, 2003].
The branching ratio can be defined as the ratio of triggered events 
over total seismicity when looking at catalog of
seismicity at large scale. The branching ratio is also equal to the
ratio of the number of secondary and later-generations aftershocks over the
total number of aftershocks within a single aftershock sequence.

\section{Formal solution of the earthquake forecast problem in the
    ETAS model}

Having stressed the importance of the indirect triggered seismicity in
determining both the overall level of seismicity and its decay law,
we now formulate the task of earthquake forecast within this model of
triggered seismicity restricted to the time domain. In this paper,
we do not address the delicate issue related to the fact that not all
earthquakes are observable or observed. Indeed, calibrations of
the ETAS parameters using the magnitude cut-offs dictated by the
requirement of seismic catalog completeness rather than by the physics of
triggered seismicity may lead to misleading results,
as unobserved events may play a significant role (in their
sheer number) in triggering observable seismicity. To our knowledge,
all previous calibrations of real seismic catalogs have bypassed this
problem, which will be studied using a technique derived from our 
renormalized Omori law in a subsequent paper.

We first summarize the single source prediction problem, which
has been studied previously by {\it Helmstetter and Sornette} [2002a].
We then consider the complete prediction problem and derive
analytically the solution for the future seismicity rate
triggered by all past event and by a constant external loading.

\subsection{Formulation of the global seismicity rate and 
renormalized Omori's law}

We define the ``bare propagator''  $\phi(t)$ of the seismicity
as the integral of (\ref{eq1}) over all magnitudes
\be
\phi(t)=\int \limits_{m_0}^{\infty}  P(m)~ \phi_{m}(t)~dm =n \Phi(t)~,
\label{phi}
\ee
which is normalized to $n$ since $\Phi(t)$ is normalized to $1$.
The meaning of the adjective ``bare'' will become clear below, when we
demonstrate that cascades of triggered events renormalize $\phi(t)$ into
an effective (``renormalized'' or ``dressed'') propagator $K(t)$.
This terminology is borrowed from statistical and condensed-matter physics
which deal with physical phenomena occurring at multiple scales in
which similar
cascades of fluctuations lead to a renormalization of ``bare'' into ``dressed''
properties when going from small to large scales.
See also [{\it Sornette and Sornette}, 1999; {\it Helmstetter and
Sornette}, 2002a] where this terminology was introduced in the present context.

The total seismicity rate $\lambda(t)$ at time $t$
is given by the sum of an ``external'' source $s(t)$
and the aftershocks triggered by all previous events
\be
\lambda(t)=s(t)+\sum_{i|t_i \leq t} \phi_{m_i}(t-t_i)~,
\label{lambda}
\ee
where $\phi_{m_i}(t-t_i)$ is defined by (\ref{eq1}).
Here, ``external'' source refers to the concept that $s(t)$
is the rate of earthquakes {\it not} triggered by other previous earthquakes.
This rate acts as a driving force ensuring that the seismicity does
not vanish and models the effect of the external tectonic forcing.

Taking the ensemble average of (\ref{lambda}) over many possible
realizations of the seismicity (or equivalently taking the mathematical
expectation), we obtain the following equation
for the first moment or statistical average $N(t)$ of $\lambda(t)$
[{\it Sornette and Sornette}, 1999; {\it Helmstetter and Sornette}, 2002a]
\be
N(t)=s(t)+\int \limits_{-\infty}^{t} ~\phi(t-\tau)~N(\tau)~ d\tau.
\label{N1}
\ee
The average seismicity rate is the solution of this self-consistent
integral equation, which embodies the fact that each event may start a
sequence of events which can themselves trigger secondary events and so on.
The cumulative effect of all the possible branching paths
of activity gives rise to the net seismic activity $N(t)$.
Expression (\ref{N1}) states that the seismic activity
at time $t$ is due to a possible external source $s(t)$ plus
the sum over all past times $\tau$ of the total previous activities $N(\tau)$
that may trigger an event at time $t$ according to the bare Omori
law $\phi(t-\tau)$.

The global rate of aftershocks including secondary
and later-generations aftershocks
triggered by a mainshock of magnitude $M$ occurring at $t=0$ is given by
$\rho(M) K(t)/n$, where the renormalized Omori law $K(t)$ is obtained as a
solution of (\ref{N1}) with the general source term $s(t)$ replaced by
the Dirac function $\delta(t)$:
\be
K(t)=\delta(t)+\int \limits_0^t~\phi(t-\tau)~K(\tau)~d\tau~.
\label{K}
\ee
The solution for $K(t)$ can be obtained as the following  series
[{\it Helmstetter and Sornette}, 2002a]
\be
K(t)=\delta(t) + {1\over 1-n} ~{{t^*}^{-\theta} \over t^{1-\theta}}~
\sum_{k=0}^{k=\infty}~(-1)^k~ {(t/t^*)^{k\theta} \over \Gamma((k+1)\theta)}~.
\label{K2}
\ee
The infinite sum expansion is
valid for $t>c$, and $t^*$ is a characteristic time measuring the distance
to the critical point $n=1$ defined by
\be
t^*=c \Bigl({n\Gamma(1-\theta) \over |1-n|} \Bigr)^{1/\theta}~.
\label{n}
\ee
$t^*$ is infinite for $n=1$ and becomes very small for $n \ll 1$.
The leading behavior of $K(t)$ at short times reads
\be
K(t)=  {1\over 1-n} ~{1 \over \Gamma(\theta)}~
{{t^*}^{-\theta} \over t^{1-\theta}}~,~~~~~~~{\rm for}~c<t<t^*~,
\label{K2bis}
\ee
showing that the effect of the cascade of secondary aftershocks renormalizes
the bare Omori law $\Phi(t) \sim 1/t^{1+\theta}$ given by (\ref{Phi})  into
$K(t) \sim  1/t^{1-\theta}$, as illustrated by Figure \ref{figKphi}.

Once the seismic response $K(t)$ to a single event is known, the complete
average seismicity rate $N(t)$ triggered by an arbitrary source
$s(t)$ can be obtained
using the theorem of Green functions for linear equations with source terms
[{\it Morse and Feshbach}, 1953]
\be
N(t)=\int \limits_{-\infty}^{t}~s(\tau)~K(t-\tau)~d\tau.
\label{N}
\ee
Expression (\ref{N}) provides the general solution of (\ref{N1}).

\subsection{The multiple source prediction problem}

We assume that seismicity which occurred in the past until the
``present'' time $u$, and which does trigger future events,
is observable. The seismic catalog consists
of a list of entries $\{(t_i, m_i), t_i<u\}$ giving
the times $t_i$ of occurrence of the earthquakes and their magnitude $m_i$.
Our goal is to set up the best possible predictor for the seismicity rate for
the future from time $u$ to time $t>u$, based on the knowledge of this catalog
$\{(t_i, m_i), t_i<u\}$. The time difference $t-u$ is called the horizon.
In the ETAS model studied here, magnitudes are determined
independently of the
seismic rate, according to the GR distribution. Therefore,
the sole meaningful target for prediction is the seismic rate. Once its
forecast is issued, the prediction of strong earthquakes is obtained
by combining the GR law with the forecasted seismic rate.

The average seismicity rate $N(t)$ at time $t>u$ in the future is
made of two contributions:
\begin{itemize}
\item the external source of seismicity of intensity $\mu$ at time $t$ plus
the external source of seismicity that occurred between $u$ and $t$
and their following aftershocks that may trigger an event at time $t$;
\item the earthquakes that have occurred in the past
at times $t_i<u$ and all the events they triggered between
$u$ and $t$ and their following aftershocks that may trigger an event
at time $t$.
\end{itemize}
We now examine each contribution in turn.

\subsubsection{Seismicity at times $t>u$ triggered by a constant source
$\mu$ active from $u$ to $t$}

Using the external seismicity source $\mu$ to forecast the
seismicity in the future would underestimate the seismicity rate because
it does not take into account the aftershocks of the external loading.
On the contrary, using the ``renormalized'' seismicity rate $\mu/(1-n)$
derived in [{\it Helmstetter and Sornette}, 2003] would overestimate the 
seismicity rate because the earthquakes that were triggered before time $u$ 
by the external source would be counted twice, since they are registered in 
the catalog up to time $u$. The correct procedure is therefore to evaluate 
the rate of seismicity triggered by a constant source $\mu$ starting at
time  $u$ to remove the influence of earthquakes that have been recorded at times 
less than $u$, whose influence for times larger than $u$ is examined later.

The response $K_{\mu}(t)$ of the seismicity to a constant source term $\mu$
starting at time $u$ is obtained using (\ref{N}) as
\be
K_{\mu,u}(t) = \mu ~\int \limits_{u^+}^t~
[K(t-\tau)-\delta(t-\tau)]~d\tau = \mu ~{\cal K}(t-u)~,
\label{Kmu}
\ee
where ${\cal K}(t)$ is the integral of $K(t)-\delta(t)$ given by (\ref{K2})
from the lower bound $u$ excluded (noted $u^+$) to $t$. This yields
\be
{\cal K}(t)	= {1 \over 1-n} ~ \Bigl({t \over t^*}\Bigr)^{\theta}
	~\sum_{k=0}^{k=\infty}(-1)^k~ {(t/t^*)^{k\theta}
	\over \Gamma((k+1)\theta+1)}~.
\label{calK}
\ee
For times larger that $t \gg t^*$, $K_\mu(t)$ reaches the asymptotic value
$K_\mu={\mu \over 1-n}$. Expression (\ref{Kmu}) takes care of both the
external source seismicity of intensity $\mu$ at time $t$ and
of its aftershocks and their following aftershocks from time $u$ to $t$
that may trigger events at time $t$.

\subsubsection{Hypermetropic renormalized propagator}

We now turn to the effect  counted from time $u$ 
of the past known events prior to time $u$
on future $t>u$ seismicity, taking into account the direct, secondary and
all later-generation
aftershocks of each earthquakes that have occurred in the past
at times $t_i<u$. Since the ETAS model is linear in the
rate variable,
we consider first the problem of a single past earthquake at time $t_i<u$ and
will then sum over all past earthquakes.

A first approach for estimating the seismicity at $t>u$ due to event $i$ that
occurred at time $t_i<u$ is to use the bare propagators $\Phi(t-t_i)$, as done
e.g. by {\it Kagan and Jackson} [2000]. This extrapolation leads to
an underestimation of the seismicity rate in the future because it does
not take into account the secondary aftershocks. This is quite a bad
approximation when $n$ is not very small, and especially for $n>0.5$, since
the secondary aftershocks are then more numerous than direct aftershocks.

An alternative would be to express the seismicity at $t>u$ due to an
earthquake that occurred at $t_i<u$ by the
global propagator $K(t-t_i)$. However, this approach would overestimate
the seismicity rate at time $t$ because of double counting. Indeed, $K(t-t_i)$
takes into account the effect of all events triggered by event $i$, including
those denoted $j$'s that occurred at times $t_i<t_j<u$ and which are directly
observable and counted in the catalog. Thus, using $K(t-t_i)$ takes into
account these events $j$, that are themselves part of the sum
of contributions over all events in the catalog.

The correct procedure is to calculate the seismicity at $t>u$ due to event $i$
by including all the seismicity that it triggered only after time
$u$. This defines what we term the ``hypermetropic renormalized propagator''
$K_u^*(t-t_i)$. It is ``renormalized'' because it takes into account secondary
and all subsequent aftershocks. It is ``hypermetropic'' because this
counting of
triggered seismicity starts only after time $u$ such that this propagator is
oblivious to all the seismicity triggered by event $i$ at short
times from $t_i$ to $u$.

We now apply these concepts to estimate the
seismicity triggered directly or indirectly by a mainshock with magnitude $M$
that  occurred in the past at time $t_i$ while
removing the influence of the triggered events
$j$ occurring between $t_i$ and $u$. This gives the rate
\be
S_M(t)= {\rho(M) \over n}~K_{u}^*(t-t_i) ~,
\label{S}
\ee
where the hypermetropic renormalized propagator $K_u^*$ is given by
\be
K_u^*(t)=\int \limits_u^t~\phi(\tau)~K(t-\tau)~d\tau~.
\label{Kstar}
\ee

$K_u^*(t)$ defined by (\ref{Kstar})
recovers the bare propagator $\Phi(t)$ for $t \approx u$,
i.e., when the rate of direct aftershocks dominates the rate of secondary
aftershocks triggered at time $t>u$. Indeed,
taking the limit of (\ref{Kstar}) for $u \to t$ gives
\ba
K_{u \to t}^*(t) &=& \int \limits_{u \to t}^t ~\phi(\tau)~K(t-\tau)~d\tau
\nonumber \\
&=&\int  \limits_{u \to t}^t ~\phi(\tau)~\delta(t-\tau)~ d\tau=\phi(t)~.
\label{lim1}
\ea
This result simply means that the forecasting of future seismicity in the near
future is mostly dominated by the sum of the direct Omori laws of all past
events. This limit recovers procedures used by {\it Kagan and Jackson} [2000].

In the other limit, $u \approx t_i$, i.e., for an event that occurred
at a time $t_i$ just before the present $u$,
$K_u^*(t)$ recovers the dressed propagator $K(t)$ (up to a
Dirac function) since there are no other
registered events
between $t_i$ and $t$ and all the seismicity triggered by event $i$
must be counted.
Using equation (\ref{K}), this gives
\be
K_{u \to 0}^*(t) = \int \limits_{u \to 0}^t ~ \phi(\tau)~K(t-\tau) ~ d\tau
	= K(t) -\delta(t)~.
\ee

Using again (\ref{K}), we can rewrite (\ref{Kstar}) as
\be
K_u^*(t)=K(t)-\int \limits_0^u K(t-\tau)~\phi(\tau)~d\tau~.
\label{Kstar2}
\ee
Putting (\ref{K2}) in (\ref{Kstar}) we obtain for $t \geq c$
\ba
K_u^*(t) &=& {\theta \over \Gamma(\theta)~\Gamma(1-\theta)}~
\int \limits_0^{t-u} ~{1 \over (t-x+c)^{1+\theta}}~{1 \over x^{1-\theta}} ~
\nonumber \\
&& \sum_{k=0}^{\infty}\Bigl [ (-1)^k~{(x/t^*)^{k\theta} \over
\Gamma((k+1)\theta) } \Bigr]~dx ~,
\label{beta2x}
\ea
where $x=t-\tau$.

We present below useful asymptotics and approximations of
$K_u^*(t)$.
\begin{itemize}
\item Hypermetropic renormalized propagator for $t \ll t^*$

Putting the asymptotic expansion of $K(t)$ for $t<t^*$ (\ref{K2bis}) in
      (\ref{Kstar}) we obtain for $t \gg c$ and $t>u$
\be
K^*_{u<t<t^*} (t) = {1 \over \Gamma(\theta)~\Gamma(1-\theta)}
~{(t+c-u)^{\theta} \over (u+c)^{\theta}~(t+c)}~,
\label{Kstartinf}
\ee
which recovers $K(t)$ for $u=0$.

\item Hypermetropic renormalized propagator for $t \gg u $

In the regime $t \gg u$, we can rewrite (\ref{Kstar2}) as
\ba
K_u^*(t) &\approx& K(t) - K(t) ~ \int_0^u ~ \phi(\tau) ~ d\tau \nonumber \\
&\approx& K(t) \Bigl(1-n \Bigl(1-{c^{\theta} \over (u+c)^{\theta}}\Bigr)\Bigr)
\label{Ks3}
\ea

\item Hypermetropic renormalized propagator for $t \approx u $

In the regime $t \approx u$ and $t-u \gg c$, we can rewrite (\ref{Kstar}) as
\ba
K_u^*(t) &\approx& \phi(t)~\int \limits_u^t~K(t-\tau)~d\tau 	\nonumber \\
&\approx&  \phi(t) ~ ( {\cal K}(t-u) +1)~,
\label{Ks4}
\ea
where ${\cal K}(t)$ is the integral of $K(t)-\delta(t)$ given by (\ref{calK}).
The additional $1$ in (\ref{Ks4}) comes from the Dirac $\delta(t)$ in
the expression
(\ref{K}) of $K(t)$.
\end{itemize}

We have performed numerical simulations of the ETAS model to test our
predictions on the hypermetropic renormalized propagator $K_u^*(t)$
(\ref{Kstar},\ref{beta2x}).

For the unrealistic case where $\alpha=0$, i.e., all events
trigger the same number of aftershocks whatever their magnitude, 
the simulations show
a very good agreement (not shown) between the results obtained by averaging over
1000 synthetic
catalogs and the theoretical prediction (\ref{Kstar},\ref{beta2x}).

Figure \ref{figKstar05} compares the numerical simulations with our
theoretical prediction for
more realistic parameters $\alpha=0.5$
with $n=1$, $c=0.001$ day, $\theta=0.2$ and $\mu=0$.
In the simulations, we construct 1,000 synthetic
catalogs, each generated by a large event that happened at time $t=0$.
The numerical hypermetropic renormalized propagator 
or seismic activity $K_u^*(t)$ is obtained by removing for each catalog
the influence of aftershocks that were triggered in the past $0<t_i<u$
where the present is taken equal to $u=10$ days and then by 
averaging over the 1,000 catalogs. It can then be compared with the
theoretical prediction (\ref{Kstar},\ref{beta2x}).
Figure \ref{figKstar05} exhibits a very good agreement 
between the realized hypermetropic seismicity rate (open circles)
and $K_u^*(t)$ predicted by (\ref{Kstar}) and shown as the continuous line up to 
times $t-u \sim 10^3 ~u$. 
The hypermetropic renormalized propagator $K_u^*(t)$ is significantly
larger than the
bare Omori law $\Phi(t)$ but smaller than the renormalized propagator $K(t)$
as expected. Note that $K_u^*(t)$ first increases with the horizon $t-u$ up to
horizons of the order of $u$ and then crosses over to a decay law
$K_u^*(t) \sim 1/t^{1-\theta}$ parallel to the dressed propagator $K(t)$.
At large times however, one can observe a new effect in the clear deviation between
our numerical average of the realized seismicity and the 
hypermetropic seismicity rate $K_u^*(t)$. This deviation pertains to 
a large deviation regime and is due to 
a combination of a survival bias and large fluctuations in the numerics.
For the larger value $\alpha=0.8$, which may be more relevant for seismicity,
the deviation between the simulated seismicity rate and the prediction
is even larger. Indeed, for $\alpha \geq 1/2$, {\it Helmstetter et
  al.} [2003] have shown that the distribution of first-generation 
seismicity rate is a power law with exponent
less than or equal to $2$, implying that its variance is ill-defined (or
mathematically infinite). Thus, averages converge slowly, all the more so
at long times where average rates are small and fluctuations are huge. 
Another way to state the problem is that, in this regime $\alpha \geq 1/2$,
the {\it average} seismicity may be a poor estimation of the {\it typical}
or {\it most probable} seismicity. Such effect can be taken into account
approximately by taking into account the coupling
between the fluctuations of the local rates and the realized magnitudes
of earthquakes at a coarse-grained
level of description [{\it Helmstetter et al.}, 2003]. 
A detailed analytical
quantification of this effect for $K(t)$ (already
discussed semi-quantitatively in [{\it Helmstetter et al.}, 2003]) and for 
$K_u^*(t)$, with an emphasis on the difference between the
average, the most probable and different quantiles of the distribution
of seismic rates, will be reported elsewhere.

Since we are aiming at predicting single catalogs, we shall resort below to
robust numerical estimations of $K(t)$ and $K_u^*(t)$ obtained by
generating numerically
many seismic catalogs based on the known seismicity up to time $u$.
Each such catalog synthesized for times $t>u$ constitutes a possible
scenario for the future seismicity. Taking the average and
calculating the median as well as different quantiles over many such scenarios
provides the relevant predictions of future seismicity
for a single typical catalog as well as its confidence intervals.

\section{Forecast tests with the ETAS model}

Because our goal is to estimate the intrinsic limits of
predictability in the ETAS model, independently of the additional
errors coming from the uncertainty in the estimation of the ETAS
parameters in real data, we consider only synthetic catalogs generated
with the ETAS model.
Testing the model directly on real seismicity would amount to test
simultaneously several hypotheses/questions: (i) ETAS is a
good model of real seismicity, (ii) the method of inversion of the
parameters is correctly implemented and stable, what is the
absolute limit of predictability of (iii) the ETAS model and (iv) of
real seismicity? Since each of these four points are difficult to address
separately and is not solved at all, we
use only synthetic data generated with the ETAS model to test the
intrinsic predictive skills of the model (iii), independently of the
other questions. Our approach thus parallels several previous attempts
to understand the degree and the limits of predictability in
models of complex systems, developed in particular as models of
earthquakes (see of instance [{\it Pepke and Carlson}, 1994;
{\it Pepke et al.}, 1994; {\it Gabrielov et al.}, 2000]).

Knowing the times $t_i$ and magnitude $m_i$ of all events that
occurred in the past up to the present $u$, the mean seismicity rate $N_u(t)$
forecasted for the future $t>u$ by taking into account all triggered events and
the external source $\mu$ is given formally by
\be
N_u(t)=\mu {\cal K}(t-u)+\sum_{i|t_i<u} {\rho(m_i) \over n} ~ K_u^*(t-t_i)~,
\label{mfmala}
\ee
where $K_u^*(t-t_i)$ is given by (\ref{Kstar}) and ${\cal K}(t)$ is given by
(\ref{calK}). In the language of the statistics of point processes, expression
(\ref{mfmala}) amounts to using
the conditional intensity function. The conditional intensity function
gives an unequivocal answer to the question of what is
the best predictor of the process. All future behaviors of the process,
starting from the present time $u$ and conditioned by the history up to time
$u$, can be simulated exactly once the form of the conditional intensity is
known. To see this, we note that the
conditional intensity function, if projected forward on the assumption
that no additional events are observed (and assuming no external variables
intervene), gives the hazard function of the time to the
next occurring event past $u$. So if the simulation proceeds by using this
form of hazard function, then by recording the time of the next event when
it does occur, and so on, ensures that one is always working with the exact
formula for the conditional distributions of the inter-event
times. The simulations then truly represent
the future of the process, and any functional can be taken from them in
whatever form is suitable for the purpose at hand.

In practice, we thus use the catalog of known earthquakes up to time
$u$ and generate many different possible scenarios for the seismicity
trajectory
which each take into account all the relevant past triggered
seismicity up to the
present $u$. For this, we use the thinning simulation method, as explained by
{\it Ogata} [1999]. We then define the average, the median and other quantiles
over these scenarios to obtain the forecasted seismicity $N_u(t)$.

\subsection{Fixed present and variable forecast horizon}

Figure \ref{predU} illustrates the problem of forecasting the aftershock
seismicity following a large $M=7$ event. Imagine that we have just
witnessed the $M=7$ event and want to forecast the seismic activity afterward
over a varying horizon from days to years in the future. In this
simulation, $u$ is kept fixed at the time just after the $M=7$ event
and $t$ is varied.
A realization of the instantaneous rate of seismic activity
(number of events per day) of a synthetic catalog is shown by the
black dots. This simulation has been performed with the parameters
$n=0.8$, $\alpha=0.8$, $b=1$, $c=0.001$ day, $m_0=3$ and $\mu=1$ event per day.
This single realization is compared with two forecasting algorithms: the sum of
the bare propagators of all past events $t_i\leq u$ (crosses), and the median
of the seismicity rate obtained over 500 scenarios generated with the
ETAS model, using the same parameters as used for generating the synthetic
catalog we want to
forecast, and taking into account the specific realization of events in each
scenario up to the present. Figure \ref{predUlog} is the same as Figure
\ref{predU} but shows the seismic activity as a function of the
logarithm of the time after the mainshock. These two figures illustrate
clearly the importance of taking into account all the cascades of
still unobserved triggered events in
order to forecast correctly the future rate of seismicity beyond a few minutes.
The aftershock activity forecast gives a very reasonable estimation of the
future activity rate, while the extrapolation of the bare Omori law
of the strong $M=7$ event together with the past seismicity very badly
under-estimates the future seismicity beyond half-an-hour after the
strong event.

\subsection{Varying ``present'' with fixed forecast horizon}

Figure \ref{Nt3005T5} compares a single realization of the seismicity rate
(open circles) observed and summed over a 5 days period and divided by $5$
so that it is expressed as a daily rate, with the predicted seismicity rate
using either the sum of the bare propagators of
the past seismicity (dots) or using the median of 100 scenarios
(crosses) generated with the same parameters as for the synthetic catalog we
want to forecast:  $n=0.8$, $c=0.001$ day, $\mu=1$ event per day,
$m_0=3$, $b=1$
and $\alpha=0.8$.
The forecasting methods calculate the total number of events over
each 5 day period
lying ahead of the present, taking into account all past seismicity
including the still unobserved triggered seismicity.
This total number of forecasted events is again divided by $5$ to express
the prediction as daily rates.
The thin solid lines indicate the first and 9$^{th}$ deciles of the
distributions of the number of events observed in the pool of 100
scenarios. Stars indicate the occurrence of large $M\geq 7$ earthquakes.
Only a small part of the whole time period used for the forecast is shown,
including the largest $M=8.5$ earthquake of the catalog,
in order to illustrate the difference between the
realized seismicity rate and the different methods of forecasting.

The realized seismicity rate is always larger than the seismicity
rate predicted using the sum of the bare propagators of the past activity.
This is because the seismicity that will occur up to 5 days in the
future is dominated by the seismicity that will be triggered in the
near future that is still unobserved but must be taken into account.
The realized seismicity  rate is close to the median of the scenarios
(crosses), and the fluctuations of the realized seismicity rate are in
good agreement with the expected fluctuations measured by the deciles
of the distributions of the seismicity rate over all scenarios generated.

Figure \ref{Nobspred3005T5} compares the predictions of
the seismicity rate over a 5 day horizon with the
seismicity of a typical synthetic catalog; a small fraction
of the history was shown in Figure \ref{Nt3005T5}.
This comparison is performed by plotting the predicted number
of events in each 5 day horizon window as a function of the actual
number of events. The open circles (crosses) correspond
to the forecasts using the median of 100 scenarios (the
sum of the bare Omori propagators of
the past seismicity). This Figure uses a synthetic catalog of $N=200000$ events
of magnitude larger than $m_0=3$ covering a time period of 150 yrs.
The dashed line corresponds to the perfect prediction when the predicted
seismicity rate is equal to the realized seismicity rate.
This Figure shows that the best predictions are obtained using the
median of the scenarios rather than using the bare propagator, which always
underestimates the realized seismicity rate, as we have already shown.

The most striking feature of Figure \ref{Nobspred3005T5} is the
existence of several
clusters, reflecting two mechanisms underlying the realized seismicity.
\begin{enumerate}
\item cluster LL with large predicted seismicity and large realized
seismicity;
\item cluster SL with small predicted seismicity and large realized
seismicity;
\item cluster SS with small predicted seismicity and small realized
seismicity;

\end{enumerate}
Cluster LL along the diagonal reflects the predictive
skill of the triggered seismicity algorithm: this is when future
seismicity is triggered by past seismicity. Cluster SL lies horizontally
at low predicted seismicity rates and shows that large seismicity rates
can also be triggered by an unforecasted strong earthquake,
which may occur even when the seismicity rate
is low. This expresses a fundamental limit of
predictability since the ETAS model permits large events
even for low prior seismicity, as the earthquake magnitudes are
drawn from the GR independently of the seismic rate.
About 20\% of the large values of the realized seismicity rate
above 10 events per day fall in the LL cluster, corresponding to a
predictability of about 20\% of the large peaks of realized seismic
activity. The cluster SS is consistent with a predictive skill
but small seismicity is not usually an interesting target.
Note there is no cluster of
large predicted seismicity associated with small realized seismicity.

Figure \ref{Nt3005T50} is the same as Figure \ref{Nt3005T5} for a
longer time window
of 50 days, which stresses the importance of taking into account
the yet unobserved future seismicity in order to accurately forecast
the level of future seismicity. Figure \ref{Nobspred3005T50} is the
same as Figure \ref{Nobspred3005T5} for the
forecast time window of 50 days with forecasts updated each 50 days.
Increasing the time window $T$ of the forecasts from 5 to 50 days leads
to a smaller variability of the predicted seismicity rate. However,
fluctuations of the seismicity rate of one order of magnitude can
still be predicted with this model. The ETAS model therefore performs
much better than a Poisson process for large horizons of 50 days.

\subsection{Error diagrams and prediction gains}

In order to quantify the predictive skills of different prediction algorithms
for the seismicity of the next five days, we use the error diagram
[{\it Molchan}, 1991; 1997; {\it Molchan and Kagan}, 1992].
The predictions are made from the present to 5 days in the future and
are updated each 0.5 day. Using a shorter time between each prediction,
or updating the prediction after each major earthquake,
will obviously improve the predictions, because large aftershocks occur
often just after the mainshock. But in practice the forecasting procedure
is limited by the time needed to estimate the location and magnitude of
an earthquake. Moreover, predictions made for very short times in advance
(a few minutes) are not very useful.

An error diagram requires the definition of a target, for instance $M \geq 6$
earthquakes, and plots the fraction of targets that were not predicted as a
function of the fraction of time occupied by the alarms (total duration
of the alarms normalized by the duration of the catalog). We define an
alarm when the predicted seismic rate is above a threshold. Recall that
in our model the seismic rate is the physical quantity that embodies completely
all the available information on past events. All targets one might
be interested in derive from the seismic rate.

Figure \ref{erdiag3005T5} presents the error diagram for $M \geq 6$ targets,
using a time window $T=5$ days to estimate the seismicity rate, and a time
$dT =0.5$ days between two updates of the predictions.
We use different prediction algorithms, either the bare propagator (dots),
the median (circles) or the mean (triangles) number of events
obtained for the 100 scenarios already generated to obtain Figures
\ref{Nt3005T5}
and \ref{Nobspred3005T5}. Each point of each curve corresponds to a different
threshold ranging from 0.1 to 1000 events per day.
The results for these three prediction algorithms are considerably
better than those obtained for a random prediction, shown as a dashed line
for reference.

Ideally, one would like the minimum numbers of failures and the
smallest possible alarm duration.
Hence, a perfect prediction corresponds to points close to the origin.
In practice, the fraction of failure to predict is 100\% without alarms
and the gain of the prediction algorithm is quantified by how fast the
fraction of failure to predict decreases from 100\% as the fraction of
alarm duration increases. Formally, the gain $G$ reported below
is defined as the ratio of the fraction of predicted targets ($=1-$
number of failures to predict) divided by the fraction of time occupied
by alarms. A completely random prediction corresponds to $G=1$.

We observe that about $50\%$ of the $M\geq6$ earthquakes can be predicted
with a small fraction of alarm duration of about $20\%$, leading to a gain
of $2.5$ for this value of the alarm duration.
The gain is significantly stronger for shorter fractions
of alarm duration: as shown in panel (b) of Figure \ref{erdiag3005T5},
25\% of the $M \geq 6$ earthquakes can be predicted
with a small fraction of alarm duration of about $2\%$, leading to a gain
of $12.5$. The origin of this good performance for only a fraction of
the targets has been discussed in relation with Figure \ref{Nobspred3005T5},
and is associated
with those events that occur in times of large seismic rate (cluster
along the diagonal in Figure \ref{Nobspred3005T5}). Panel (c)
of Figure \ref{erdiag3005T5} shows the dependence of the prediction gain $G$
as a function of the alarm duration: the three prediction schemes give
approximately the same power law increase with an exponent
close to $1/2$ of the gain
as the duration of alarm decreases. For small alarm duration, the gain
reaches values of several hundreds. The saturation at very small values
of the alarm duration is due to the effect that only a few targets are sampled.
Figures \ref{erdiag3005T5M55} and \ref{erdiag3005T5M7} are similar to
Figure \ref{erdiag3005T5}, respectively for a smaller target of magnitudes
larger than $5$ and a larger target of magnitudes larger than $7$.

Table \ref{tabG} presents the results for the prediction gain and for the
number of successes using different choices of the time window $T$ and of
the update time $dT$ between two predictions, and for different values of
the target magnitude between 5 and 7.
The prediction gain decreases if the time between two updates
of the prediction increases, because most large earthquakes occur
at very short times after a previous large earthquake.
In contrast, the prediction gains do not depend on the time window $T$
for the same value of the update time $dT$.

The prediction gain is observed to increase significantly with the
target magnitude,
especially in the range of small fraction of alarm durations
(see Table \ref{tabG} and Figures \ref{erdiag3005T5}-\ref{erdiag3005T5M7}).
However, this increase of the prediction gain does not mean that large
earthquakes are more predictable than smaller ones, in contrast with, for
example, the critical earthquake theory [{\it Sornette and Sammis}, 1995;
{\it Jaum\'e and Sykes}, 1999; {\it Sammis and Sornette}, 2002].
In the ETAS model, the increase of the prediction gain with the
target magnitude is due to the
decrease of the number of target events with the target magnitude.
Indeed, choosing $N$ events at random in the catalog independently
of their magnitude gives on average
the same prediction gain as for the $N$ largest events. This demonstrates
that the larger predictability of large earthquakes is solely a size effect.

We now clarify the statistical origin of this size effect.
Let us consider a catalog of total duration $D$ with a
total number $N$ of events analyzed with $D/T$
time windows with horizon $T$. These $D/T$ windows can be
sorted by decreasing seismicity $r_1 > r_2 > ... > r_n > ...$,
where $r_i$ is the $i$-th largest number of events in a window of size $T$.
There are $n_1, n_2, ..., n_i, ...$ windows of type $1, 2, ..., i,
...$ respectively,
such that $\sum_i r_i~n_i=N$.
Then, the frequency-probability that an earthquake drawn at random
from the catalog falls
within a window of type $i$ is
\be
p_i = {r_i ~n_i \over N}~.
\ee
We have found that, over more than three decades spanning from $1$
event to $10^3-10^4$ events
per window, the cumulative distribution of these $p_i$'s is a power
law with exponent
approximately equal to $\kappa=0.4$. This power law
is found for the observable realized seismicity
as well as for the seismicity predicted by the different methods
discussed above.
Such a small exponent $\kappa$ implies that the few windows that happen to have
the largest number of events contain a significant fraction of the
total seismicity.
It can be shown [{\it Feller}, 1971] that, in absence of any
constraint, the single window with the largest
number of events contains on average $1-\kappa=60\%$ of the total
seismicity. This would imply that
there is a $60\%$ probability that an earthquake drawn at random
within the catalog (of $150$ years)
belongs to this single window of $5$ days. This effect is moderated
  because the random variables $p_i$ have to
sum up to $1$ by definition. This can be shown to entail a roll-off
of the cumulative distribution depleting the largest values of $p_i$.
Empirically, we find that the most active window out of the
approximately $54,000$ daily windows of our $150$ years long catalog
contains only $3\%$ of the total number of events. While this value of
$3\%$ is smaller than the prediction of 60\% in absence of
normalization, it is  considerably larger than the ``democratic''
result which would predict a fraction of about $0.002\%$ of the
seismicity in each window. Since a high seismicity rate implies 
strong interactions and triggering between earthquakes and is usually 
associated with recent past high
seismicity; the events in such a window are highly predictable.
When the number of targets increases, one starts to sample
statistically other windows with smaller seismicity which have
a weaker relation with triggered seismicity and thus present
less predictive power.

In our previous discussion, we have not distinguished the skills of the
three algorithms, because they perform essentially identically with
respect to the assigned targets. This is very surprising from
the perspective offered by all our previous analysis showing that the
naive use of the direct Omori law without taking into account
the effect of any indirect triggered seismicity strongly underestimate the
future seismicity. We should thus expect a priori that this prediction scheme
should be significantly worse than the two others based on a correct
counting of all unobserved triggered seismicity. The explanation for
this paradox is given by examining Figure \ref{dsg}, which presents
further insight in the prediction methods applied to the
synthetic catalogs used in Figures \ref{predU}-\ref{erdiag3005T5M7}.
Figure \ref{dsg} shows three quantities as a function of the threshold
in seismicity rate used to define an alarm, for each of the three algorithms.
These quantities are respectively the duration of alarms normalized
by the total
duration of the catalog shown in panel (a), the fraction of successes
($=1-$ failure to predict) shown in panel (b) and the prediction gain shown in
panel (c). These three panels tell us that the incorrect level of seismic
activity predicted by the bare Omori law approach
can be compensated by the use of a lower alarm threshold. In other words,
even if the seismicity rate predicted by the bare Omori law approach
is wrong in absolute values, its time evolution in relative terms
contains basically the same information as the full-fledged method taking
into account all unobserved triggered seismicity. Therefore, an algorithm
that can  detect a relative change of seismicity can perform as well as the
complete approach for the forecast of the assigned targets.
This is an illustration of the fact that predictions of
different targets can have very different skills which depend on the
targets. Using the full-fledged renormalized
approach is the correct and only method to get the best possible predictor of
future seismicity rate. However, other simpler and more naive methods
can perform almost as well for more restricted targets, such as the
prediction of only strong earthquakes.

\subsection{Optimization of earthquake prediction}

The optimization of the prediction method requires the definition of a
loss function $\gamma$, which should be minimized in order to
determine the optimum alarm threshold of the prediction method.
The probability gain defined in the previous section cannot be used
as an optimization method, because it is maximum for zero alarm time.
The strategies that optimize the probability gain are thus very impractical.
An error function commonly used [e.g., {\it Molchan and Kagan}, 1992]
is the sum of the two types of errors in earthquake prediction,
the fraction of alarm duration $\tau$ and the fraction of missed
events $\nu$.
This loss function is illustrated in Figure \ref{dsgd} for the same
numerical simulation of the ETAS model as in the previous
section, using a prediction time window of 5 days and an update time
of 0.5 day, for different values of the target magnitude between 5 and 7.
For a prediction algorithm that has no predictive skill, the loss
function $\gamma$ should be close to 1, independently of the alarm duration
$\tau$. This is indeed what we observe for the prediction algorithm
using a Poisson process, with a constant seismicity rate equal to the
average seismicity rate of the realized simulation.
For the prediction methods based on the ETAS model, we obtain a
significant predictability by comparison with the Poisson process
(Figure \ref{dsgd}). The loss function is minimum for a fraction of
alarm duration of about 10\%, and decreases with the target magnitude
$M_t$ from 0.9 for $M_t=5$ down to 0.7 for $M_t=7$.
We expect that the minimum loss function will be much lower when using
the information on earthquake locations.

\subsection{Influence of the ETAS parameters on the predictability}

We now test the influence of the model parameters 
$\alpha$, $n$, $m_0$ and $p$ on the predictability.
We did not test the influence of the $b$-value, because this parameter
is rather well constrained in seismicity, and because its influence
is felt only relative to $\alpha$. We keep $c$ equal to 0.001 day
because this parameter is not critical as long as it is small. The
external loading $\mu$ is also fixed to 1 event/day because it is 
only a multiplicative factor of the global seismicity.
The value of the minimum magnitude $m_0$ above which earthquakes may 
trigger aftershocks of magnitude larger than $m_0$ is very poorly constrained.
This value is no larger than $3$ for the Southern California 
seismicity, because there are direct evidences of $M3+$ earthquakes triggered 
by $M=3$ earthquakes [{\it Helmstetter}, 2003].
We are limited in our exploration of small
values of $m_0$ because the number of earthquakes
increases rapidly if $m_0$ decreases, and thus the computation time
becomes prohibitive. We have tested the values $m_0=1$ and $m_0=2$ and
find that the effect of a decrease of $m_0$
is simply to multiply the seismicity rate obtained for $M\geq3$ by a constant
factor. Using a smaller $m_0$
does not change the temporal distribution of seismicity and 
therefore does not change the predictability of the system.

We use the minimum value of the error function $\gamma$ introduced in 
the previous section to characterize the predictability of the
system. This function is estimated from the seismicity rate predicted
using the bare propagator, as done in previous studies
[{\it Kagan and Knopoff}, 1987; {\it Kagan and Jackson}, 2000].
While all three prediction methods that we have investigated
give approximately the same results for
the error diagram, the method of the bare propagator is much
faster, which justifies to use it. The results are summarized in Table 
\ref{tabgamma}, which gives
the minimum value of the error function $\gamma$ for each set
of parameters, and the corresponding values of the alarm duration,
the proportion of predicted $M\geq6$ events and the prediction gain.
All values of $\gamma$ are in the range 0.6-0.9, corresponding to 
a small but significant predictability of the ETAS model.
The predictability increases (i.e, $\gamma$ decreases) with 
$\alpha$, $n$ and  $p$, because for large $\alpha$, large $n$ 
and/or large $p$, there are larger fluctuations of the seismicity 
rate which have stronger impacts for the future seismicity.
The minimum magnitude $m_0$ has no significant influence
on the predictability. We recover the same pattern when
using another estimate of the predictability measured by
the prediction gain $G_{1\%}$ for an alarm duration of $1\%$.

\subsection{Information gain}

We now follow {\it Kagan and Knopoff} [1977], who introduced the 
entropy/information concept linking the likelihood gain to the 
entropy per event and hence to the predictive power of the fitted model, and
{\it Vere-Jones} [1998], who suggested using the information gain to
compare different models and to estimate the predictability of a process.

Our forecasting algorithm provides  the average seismicity rate
$\lambda_i$ above $m_0$ in the time interval  $(t_i,t_i+T)$.
Assuming a constant magnitude distribution given by (\ref{GRlaw}),
the probability $p_i$ to have at least one event above the target
magnitude $M_t$ in the time interval $(t_i,t_i+T)$ can be evaluated
from the average seismicity rate  $\lambda_i$ by
\be
p_i=1-\exp\Bigl(-\lambda_i ~10^{-b(M_t-m_0)}\Bigr)~.
\label{pi}
\ee
Figure \ref{pimed} shows the probability $p_i$
obtained for different choices of the target magnitude $M_t$ for the same
sequence as in Figure \ref{Nt3005T5}.

The binomial score $B$  compares the prediction $p_i$ with the realization
$X_i$, with $X_i=1$ if a target event occurred in the interval
$(t_i,t_i+T)$ and $X_i=0$ otherwise.
For the whole sequence of intervals  $(t_i,t_i+T)$, the binomial score
is defined by
\be
B=\sum_i \Bigl [  X_i \log(p_i) +(1-X_i) \log(1-p_i) \Bigr ]~,
\label{B}
\ee
where the sum is performed over all (non-overlapping) time windows covering
the whole duration of the catalog. The first term represents the contribution 
to the score from those intervals which contain an event, and the second term 
the contribution to the score from those intervals which contain no event.

In order to test the performance of a forecasting algorithm, we compare
the binomial score $B$ of the forecast with the binomial score of a Poisson 
process. We use two definitions of the Poisson process.
First, we use the same definition as in [{\it Vere-Jones}, 1998] and 
take a Poisson process with a seismicity rate equal to the average seismicity 
rate of the realized catalog and use equation (\ref{pi}) to estimate
the probability of having at least an event above the target magnitude
in each time interval. Because this definition of the probability assumes
a uniform temporal distribution of target events and thus neglects
clustering of large events, it over-estimates the proportion of intervals 
which have at least one target event. Indeed, the probability of having 
several events of magnitude $M\geq M_t$  in the same time window is
much higher for the ETAS model than for a Poisson process.
We thus propose another definition of the Poisson process defined
by putting all values of $p_i$ equal to the fraction of intervals 
in the realized catalog that have at least a target events.
For small time intervals and/or large target magnitudes, the
proportion of intervals that have several target events is very small,
and the two definitions of the Poisson process give similar results.
The results for different choices of the time interval $T$ and of the
target magnitude $M_t$ are listed in Table \ref{tabB}.

We evaluate the binomial score $B$ for different prediction algorithms:
(i) the average of the seismicity rate over all scenarios ($B_{mean}$), 
(ii) the exponential of the average of the logarithm of the mean of the
seismicity rate ($B_{meanl}$), (iii) the median of the seismicity rate 
($B_{med}$) and (iv) the sum of the bare propagators of the past
seismicity ($B_{\phi}$). The results for the forecasting methods based on 
the median seismicity rate ($B_{med}$ and $B_{meanl}$) are in general 
better than the Poisson process, i.e., the binomial score for the forecasting 
algorithms are larger than the score obtained with a Poisson process 
for both definitions of the Poisson process.
The binomial score $B_{pois2}$ measured using the second definition
of the Poisson process is higher than the binomial score  $B_{pois1}$ 
for the first definition of the Poisson process because the first
method is biased and over-estimates the  probability of having at
least a target event.   
The scores of the forecasting methods which take into account the
cascade of  secondary aftershocks ($B_{med}$, $B_{mean}$ and $B_{meanl}$
in Table \ref{tabB}) are significantly better than the score  $B_{\phi}$ obtained 
with the bare propagator, even for short time intervals $T$, because
this predictor under-estimates the realized seismicity rate.
Similarly, the score $B_{mean}$ obtained for the average seismicity
rate is generally smaller than $B_{med}$ because the averaging method
generally over-estimates the realized seismicity rate. For large 
time intervals $T \geq 10$ days, and for large target magnitudes, the results
for the bare propagator are even worse than the results obtained with
a Poisson process using both definitions of the Poisson process. 

Our results are in disagreement with those reported in {\it
Vere-Jones} [1998] on the same ETAS model: we conclude that the ETAS
model has a significantly higher predictive power than the Poisson process
while {\it Vere-Jones} [1998] concludes that the forecasting performance
of the ETAS model is worse than with the Poisson model. 
Vere-Jones and Zhuang's procedure and ours are very similar.
They use the same method to generate ETAS simulations and to update
the predictions at rigid time intervals, with a similar time between
two updates of the predictions. They use the same method to estimate
the probability of having a target event for the Poisson process
(corresponding to our first definition of the Poisson process) but 
a different method to estimate the probability $p_i$ for the ETAS
model. Rather than using eq. (\ref{pi}) to derive the probability $p_i$
from the seismicity rate as done in this work, they measure directly 
this probability from the fraction of scenarios which have at least 
a target event. We have compared the two methods and found that the
method of Vere-Jones is very similar to the results derived from the 
median seismicity rate. However, for large target events and small 
time intervals, the method of Vere-Jones requires to generate a huge number of
scenarios to obtain accurate estimates of the fraction of scenarios
which have at least a target event. We thus believe that our method
might be better in this case and give a more accurate estimate of the 
probability $p_i$ of  having at least a target event. This is one
possible origin of the discrepancy between Vere-Jones' results and ours.

The ETAS parameters used in the two studies are also very similar and
cannot account for the disagreement.
The tests of {\it Vere-Jones} [1998] have been performed using a
$\alpha/b$ ratio of $0.57/1.14=0.5$ smaller than the value
$\alpha/b=0.8$ used in our
simulations. This difference may lead to a smaller predictability for the
simulations of {\it Vere-Jones} [1998] because there are fewer large
aftershock sequences. The branching ratio $n=0.78$ used by {\it
Vere-Jones} [1998] is very close to our value $n=0.8$.
However, these difference in the ETAS parameters cannot
explain why {\it Vere-Jones} [1998] obtains a better predictability
for a Poisson process than for the ETAS model. {\it Vere-Jones} [1998] 
concludes that the Poisson process is better predictive than
ETAS because the binomial score measured for time periods containing
target events, i.e., by taking only the first term of eq. (\ref{B}),
is larger for the Poisson process than for ETAS.
While we agree with this point, we note that the fact that the first
term of the binomial score is larger for the Poisson process than for
ETAS does not imply that the Poisson process is more predictive.
Indeed, the score for periods containing events is maximum
for a simple model specifying a probability $p_i=1$ of having an event 
for all time intervals. Because this simple model does not 
miss any event, it gives the maximum binomial score for periods
containing target events, but is of course not better than the ETAS or the 
Poisson model if we take into account the second term of the binomial
score corresponding to periods without target events.
If we take our second definition of the Poisson process, taking into
account clustering, we obtain a better score with the ETAS model for 
periods containing target events than for the Poisson process.
We thus think that the conclusion of {\it Vere-Jone} [1998] that the
ETAS model is sometimes less predictive than the Poisson process is 
due to an inadequate measure of the predictability.
We thus caution that a suitable assessment of the forecasting skills
of a model requires several complementary quantitative measures, such
as the predicted versus realized seismicity rates, the error diagram
and predictability gain and the entropy-information gains,
and a large number of target events.

\section{Conclusions}
\label{conclu}

Using a simple model of triggered seismicity, the ETAS model, based on
the (bare) Omori law, the Gutenberg-Richter law and the idea that large
events trigger more numerous aftershocks, 
we have developed an analytical approach to account for the
triggered seismicity adapted to the problem of forecasting future
seismic rates at varying horizons from the present. Tests 
presented on synthetic catalogs have validated the use of interacting 
triggered seismicity to forecast large earthquakes in these models.
This work provides what we believe is a useful benchmark from which
to develop real prediction tests of real catalogs. These tests have also
delineated the fundamental limits underlying forecasting skills, stemming from
an intrinsic stochastic component in the seismicity models.
Our results offer a rationale for the fact that pattern recognition algorithms
may perform better for strong earthquakes than for weaker events.
Although the predictability of an earthquake is independent of its magnitude
in the ETAS model, the prediction gain is better for the largest events
because they are less numerous and it is thus more probable that they are
associated with periods of large seismicity rates, which are themselves
more predictable.

We have shown in {\it Helmstetter et al.} [2003] that most precursory
patterns used in prediction algorithms, such as a decrease of $b$-value or
an increase of seismic activity can be reproduced by the ETAS model.
If the physics of triggering is fully characterized by the class
of models discussed here, this suggests that patterns and
precursory indicators are sub-optimal compared with the prediction based on
a full modeling of the seismicity. The calibration of the
ETAS model or some of its variants
on real catalogs as done in
{\it Kagan and Knopoff} [1987], {\it Kagan and Jackson} [2000],
{\it Console and Murru} [2001], {\it Ogata} [1988, 1989, 1992, 1999, 2001],
{\it Kagan} [1991], {\it Felzer et al.} [2002] represent important steps
in this direction.
However, in practical terms, the issue of the model errors
associated with the use of an incorrect model calibrated
on an incomplete data set with not fully known parameters
may make this statement weaker or even turn it on its head.

\acknowledgments
We are very grateful to D. Harte, Y.Y. Kagan, Y. Ogata, F. Schoenberg,
D. Vere-Jones and J. Zhuang for useful exchanges and for their feedback on
the manuscript. This work is partially supported by 
by NSF-EAR02-30429, by
the Southern California Earthquake Center (SCEC) and by
the James S. Mc Donnell Foundation
21st century scientist award/studying complex system.

\end{article}

\clearpage

\begin{figure}
\psfig{file=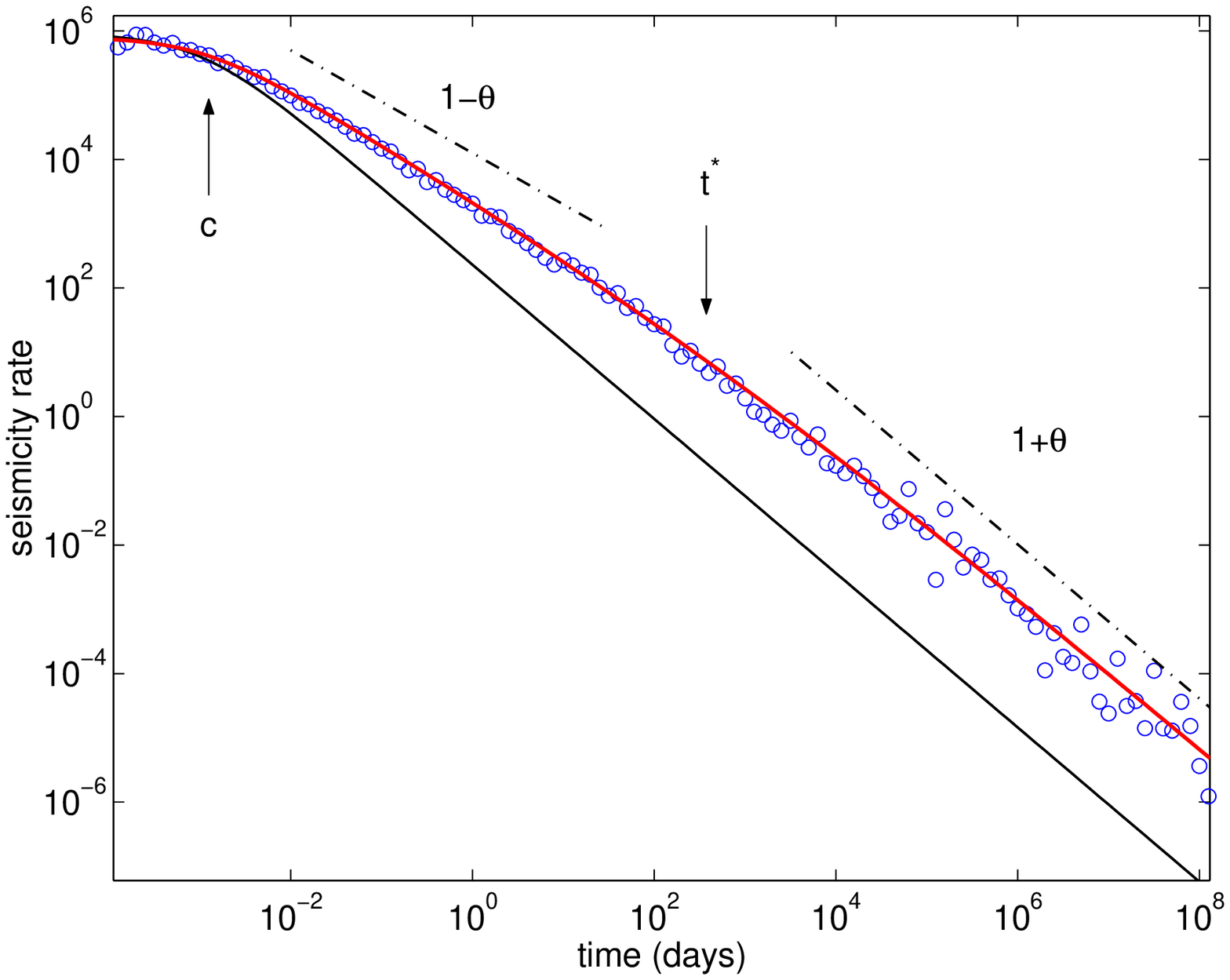,width=14cm}
\caption{\label{figKphi}
A realization of the ETAS model shows the seismicity rate (open circles)
as a function of time after a large earthquake. This
illustrates the differences between the realized seismicity rate $\lambda(t)$
(open circle), the average renormalized (or dressed) propagator
$K(t)$ (gray line), and the local propagator $\Phi_{m}(t)$ (thin black line) .
This aftershock sequence has been generated using the ETAS model with
parameters $n=0.91$, $\alpha=0.5$, $b=1$, $\theta=0.2$, $m_0=0$ and $c=0.001$
day, starting from a mainshock of magnitude $M=8$ at time $t=0$.
The global aftershock rate is significantly higher than the direct (or first
generation) aftershock rate, described by the local propagator $\Phi_m(t)$.
The value of the branching ratio $n=0.915$ implies that about $91.5$\%
of aftershocks are triggered indirectly by the mainshock.
The global aftershock rate $N(t)$ decreases on average according to the dressed
       propagator $K(t) \sim1/t^{1-\theta}$ for $t<t^*$, which is
significantly slower than
the local propagator $\phi(t) \sim 1/t^{1+\theta}$. These two asymptotic
power laws are shown as dotted-dashed lines with their slopes in the log-log
plot corresponding respectively to the exponents $1-\theta$ and $1+\theta$.
}
\end{figure}

\clearpage

\begin{figure}
\psfig{file=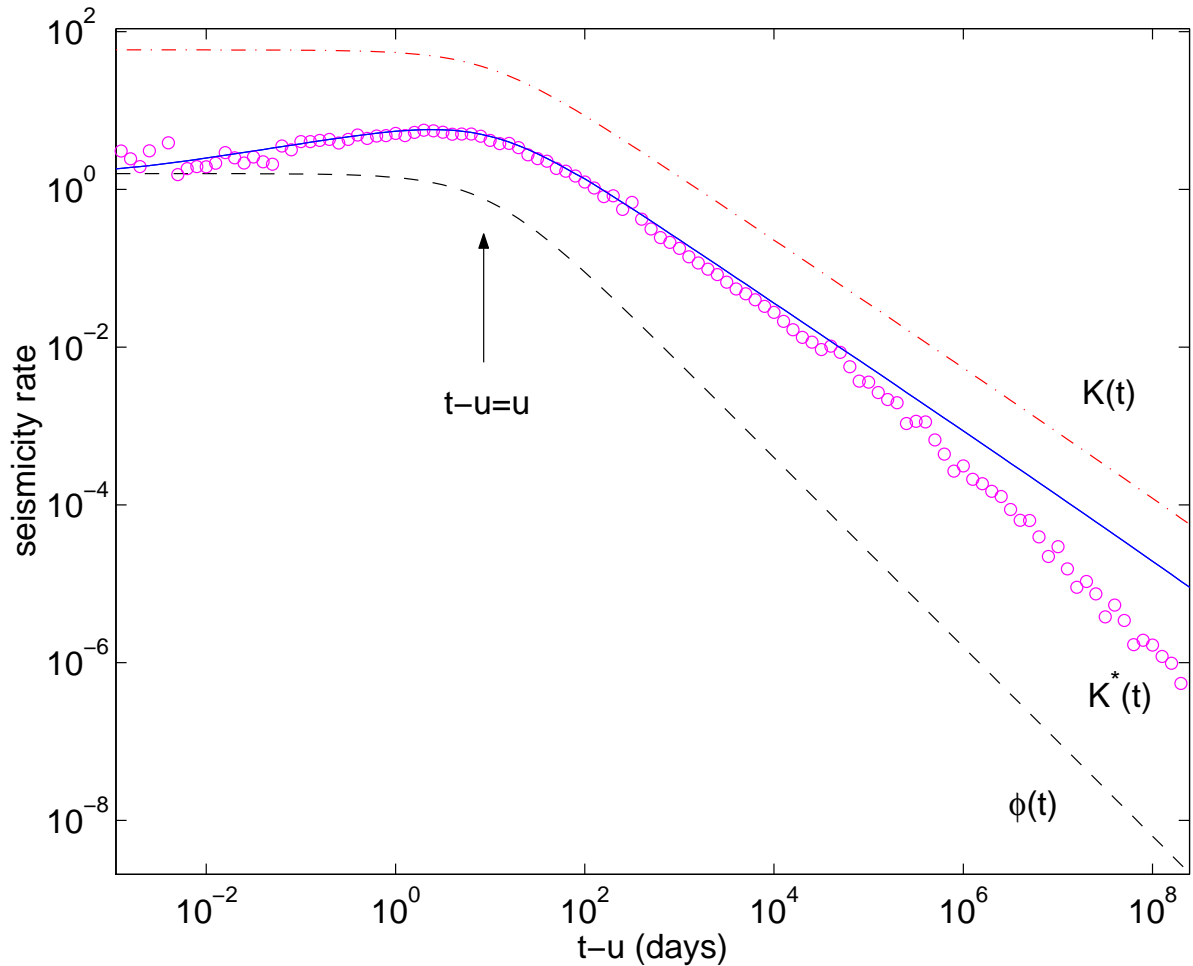,width=16cm}
\caption{\label{figKstar05}
Seismicity rate (hypermetropic renormalized propagator)
following a large event that happened at time $t=0$, removing
the influence of aftershocks that were triggered in the past
$0<t_i<u$ with $u=10$ days.
We have averaged over 1000 simulations starting at time $u=10$ days
after a large event
of magnitude $m=6$, using the parameters $n=1$, $c=0.001$ day, 
$\theta=0.2$ and $\alpha=0.5$. There is a very good agreement 
between the realized hypermetropic seismicity rate (open circles)
and $K_u^*(t)$ predicted by (\ref{Kstar}) and shown as the continuous line up to 
times $t-u \sim 10^3 ~u$. At large times, there is clear deviation between
our numerical average of realized seismicity and the 
hypermetropic seismicity rate $K_u^*(t)$ due to 
a combination of a survival bias and large fluctuations in the numerics.}
\end{figure}

\clearpage

\begin{figure}
\psfig{file=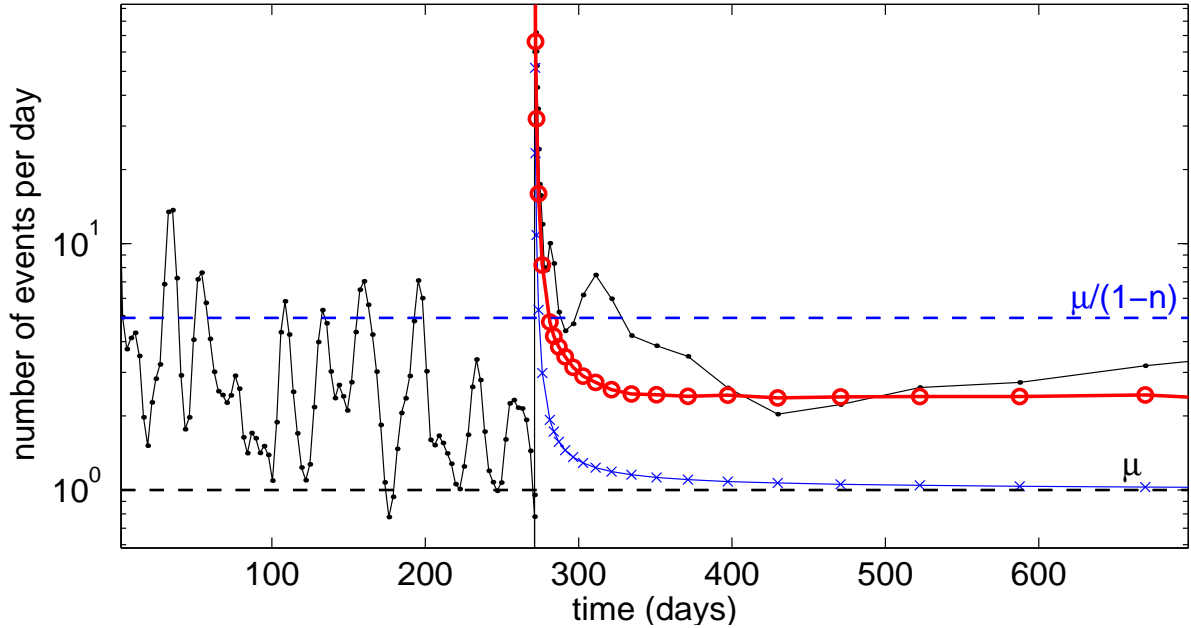,width=16cm}
\caption{\label{predU}
Rate of seismic activity for a synthetic catalog
(black dots) generated with the parameters
$n=0.8$, $\alpha=0.8$, $b=1$, $c=0.001$ day, $m_0=3$ and $\mu=1$ event per day.
We compare different methods of prediction of the seismicity rate following
a large event $M=7$ that has occurred at the time of the large peak
shown in the figure. Using the data up to
the present time $t_i\leq u$, where $u$ is the ``present'' taken
fixed just after the $M=7$ earthquake, we try to predict the
future activity
up to 1 year in the future.
We use two predictions algorithms: the sum of the bare propagators of all
       past events $t_i\leq u$ (crosses), and the median of the seismicity
rate (open circles) obtained
over 500 scenarios generated with the ETAS model, using the same
parameters
as for the synthetic catalog we want to predict, and taking into
account the specific
realization of the synthetic catalog up to the present.}
\end{figure}


\begin{figure}
\psfig{file=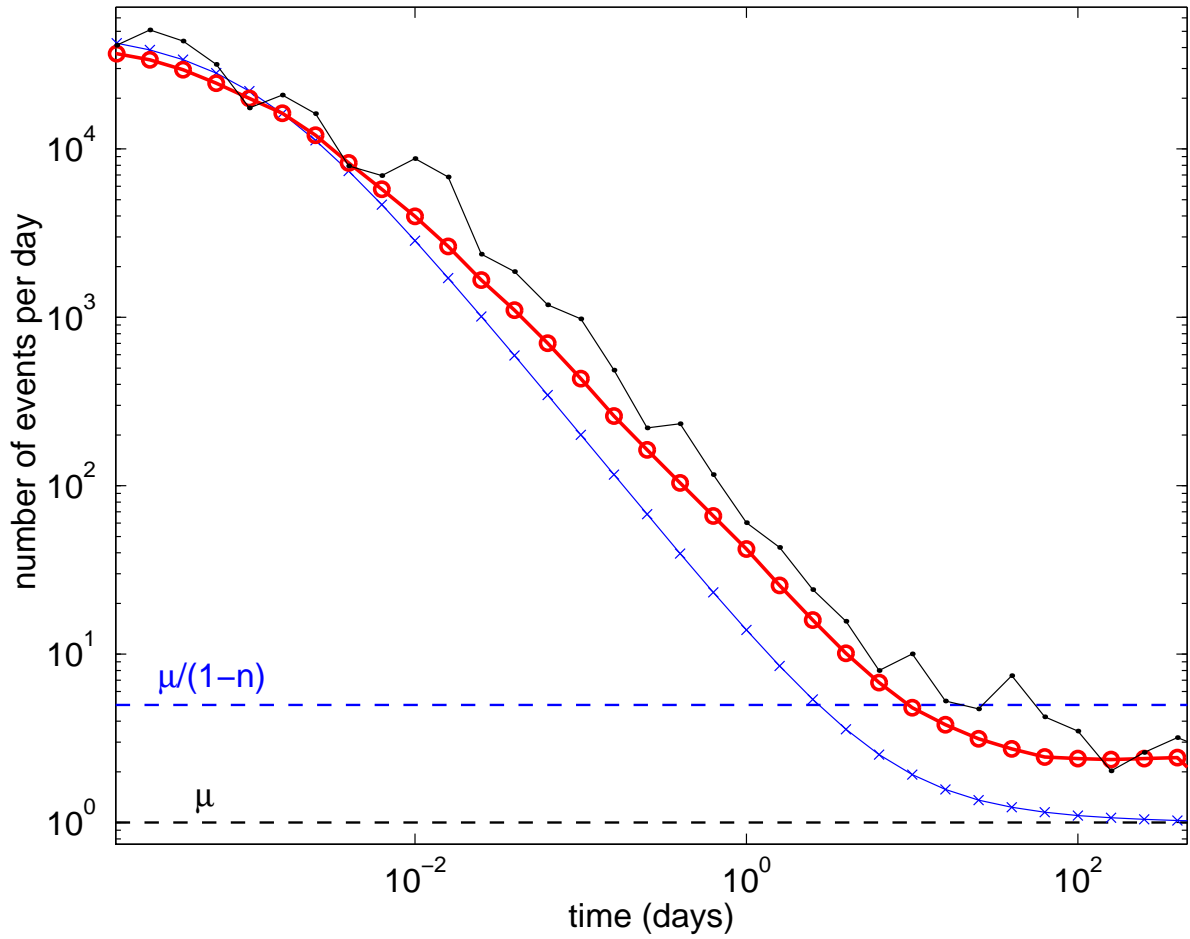,width=16cm}
\caption{\label{predUlog} Same as Figure \ref{predU} but as a function of the
logarithm of the time after the mainshock. At early time $t-u < 10^{-2}$ days,
the predicted seismicity rate is correctly predicted by the naive
bare Omori law shown by the crosses which is indistinguishable from the
more elaborate scheme involving all cascades of triggered events.
At larger times, the cascade of
triggered seismicity renormalizes the seismicity rate to a significantly higher
level, which is correctly forecasted by the mean over the 500 scenarios.
}
\end{figure}

\clearpage

\begin{figure}
\psfig{file=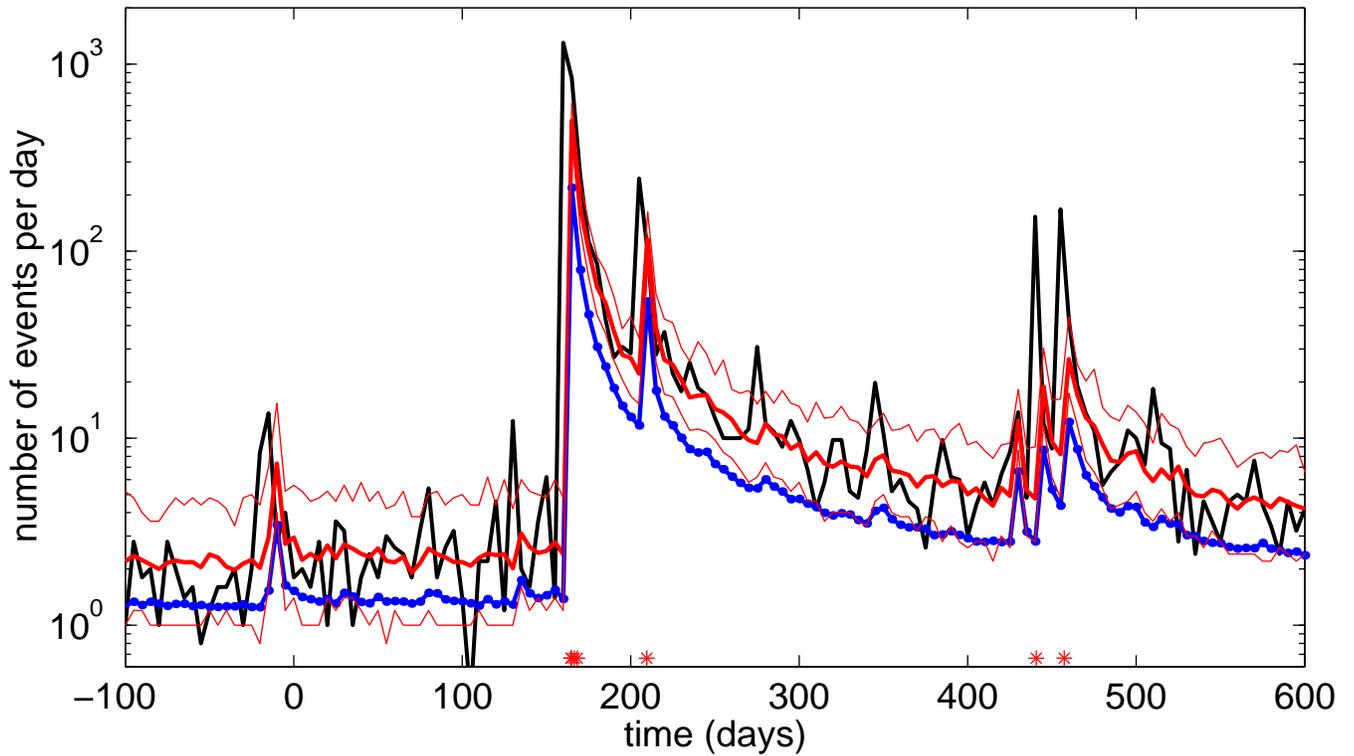,width=18cm}
\caption{\label{Nt3005T5} Comparison between the seismicity rate
(black line) observed for 5 day time periods, with the predicted
seismicity rate using either the sum of the bare propagators of the
past seismicity (dots) or using the median
of 100 scenarios (gray line) generated with the same parameters as for
the synthetic catalog we want to predict,  $n=0.8$, $c=0.001$ day, 
$\mu=1$ event per day, $m_0=3$,
$b=1$ and $\alpha=0.8$. The thin solid lines indicate the first and
9$^{th}$ deciles
of the set of 100 scenarios: there is 80\% probability that the
predicted seismicity
over the next 5 days falls within these two lines.
Stars indicate the occurrence of a large $M\geq 7$ earthquake. Forecasts
are updated every 5 days.
}
\end{figure}

\clearpage

\begin{figure}
\psfig{file=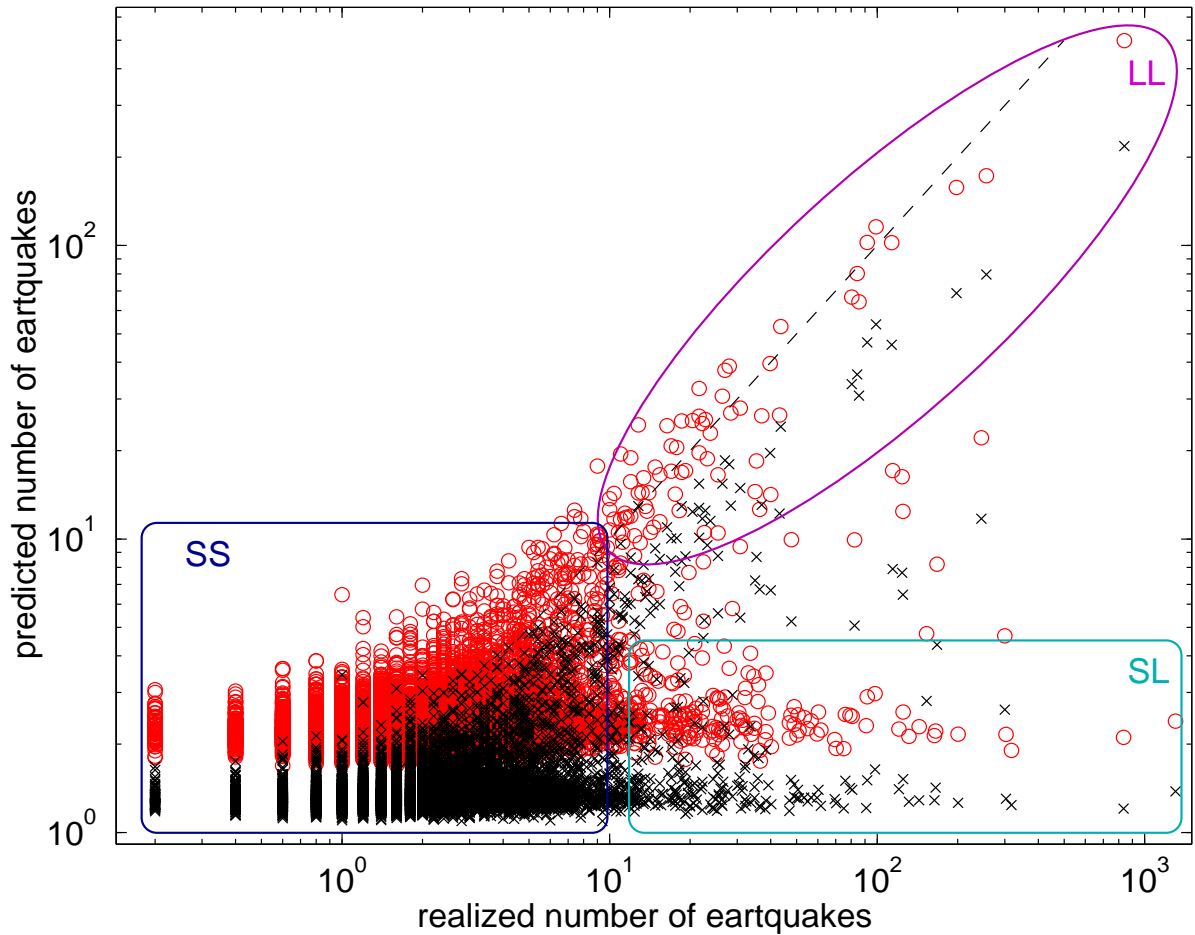,width=16cm}
\caption{\label{Nobspred3005T5}
Comparison between the seismicity rate observed
for 5 day time periods from the present with the predicted
seismicity rate over the same 5 day periods using either the sum
of the bare propagators of the past seismicity (crosses) or using the median
of 100 scenarios (circles), corresponding to the same data shown in Figure
\ref{Nt3005T5} but using a long synthetic catalog of $N=200000$ events over
a time period of 150 yrs. The dashed line corresponds to the
perfect forecast when the predicted seismicity
rate is equal to the realized seismicity rate.
This figure shows that the best forecasts are obtained using the
median of the scenarios rather than using the  bare propagator,
which always underestimates the realized seismicity rate. The meaning
of the clusters LL, SL and SS are discussed in the text.
Forecasts are updated every 5 days. A faster rate of updating does not change
the fraction of predictable events lying close to the diagonal.
}
\end{figure}

\clearpage

\begin{figure}
\psfig{file=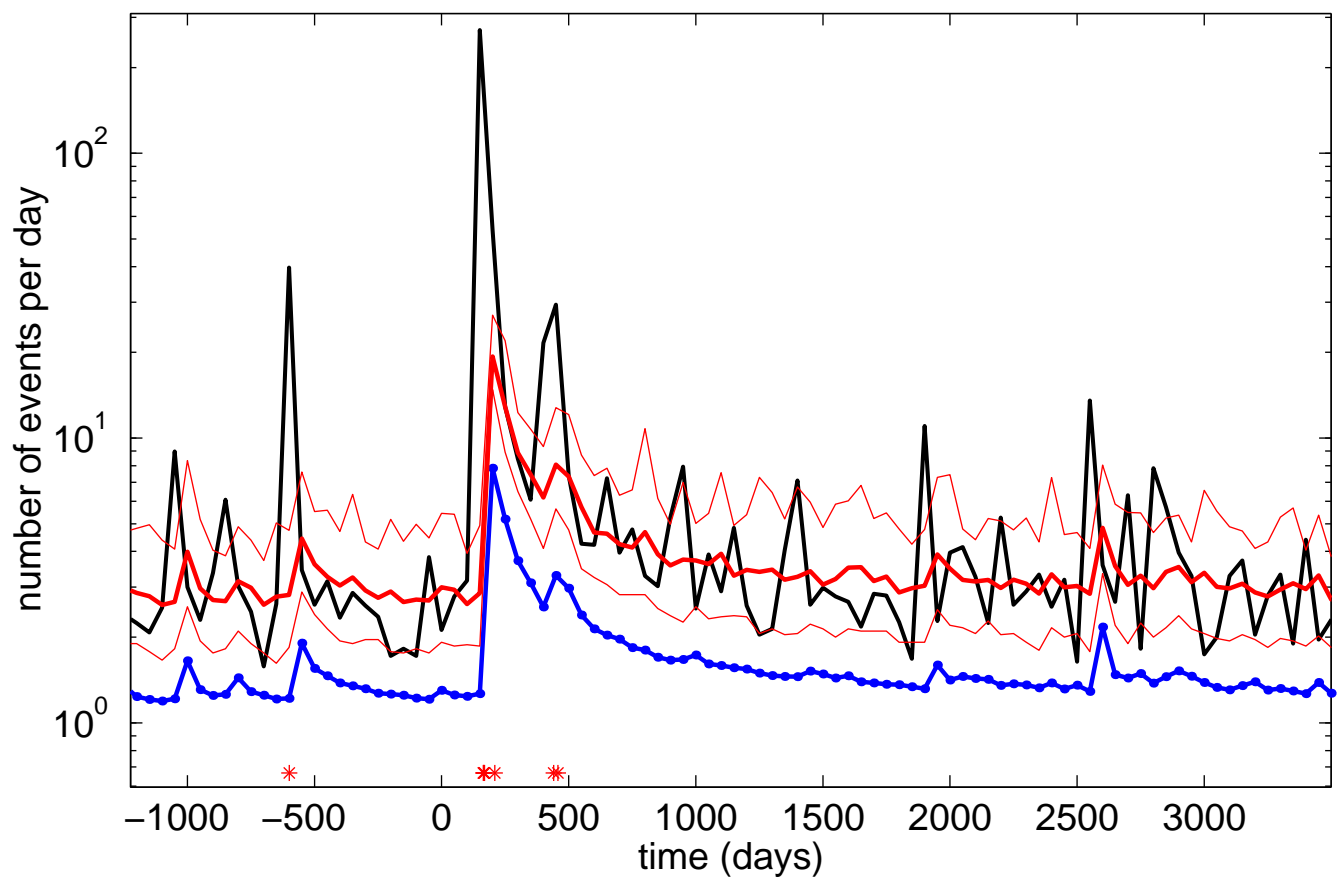,width=18cm}
\caption{\label{Nt3005T50} Same as Figure \ref{Nt3005T5}
but for a larger horizon $t=50$ days.}
\end{figure}

\clearpage

\begin{figure}
\psfig{file=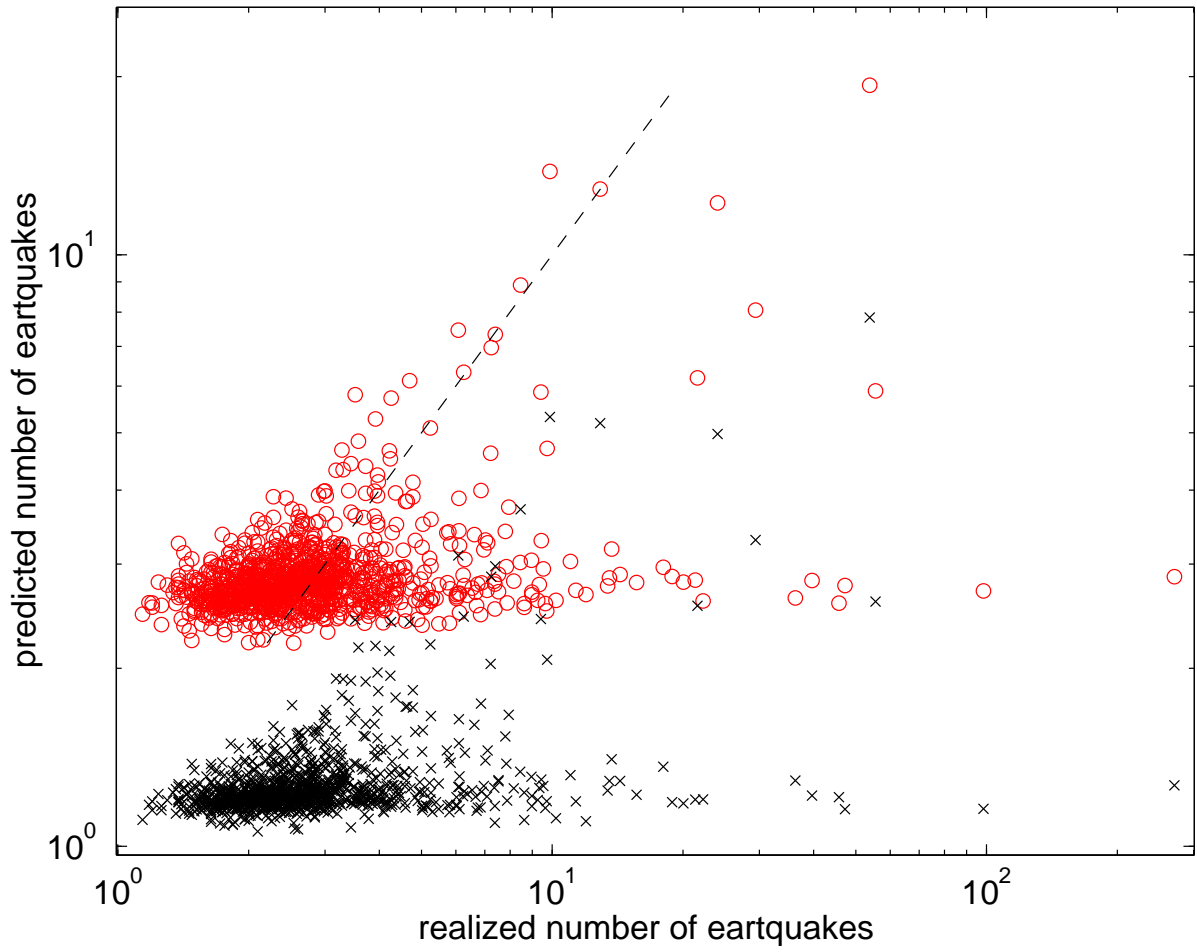,width=16cm} 
\caption{\label{Nobspred3005T50} Same as Figure \ref{Nobspred3005T5}
but for a larger horizon $t=50$ days. For such large horizons,
taking into account the cascade of triggered seismicity makes a large
difference on the performance of the predicted seismicity rate.
}
\end{figure}

\clearpage

\begin{figure}
\psfig{file=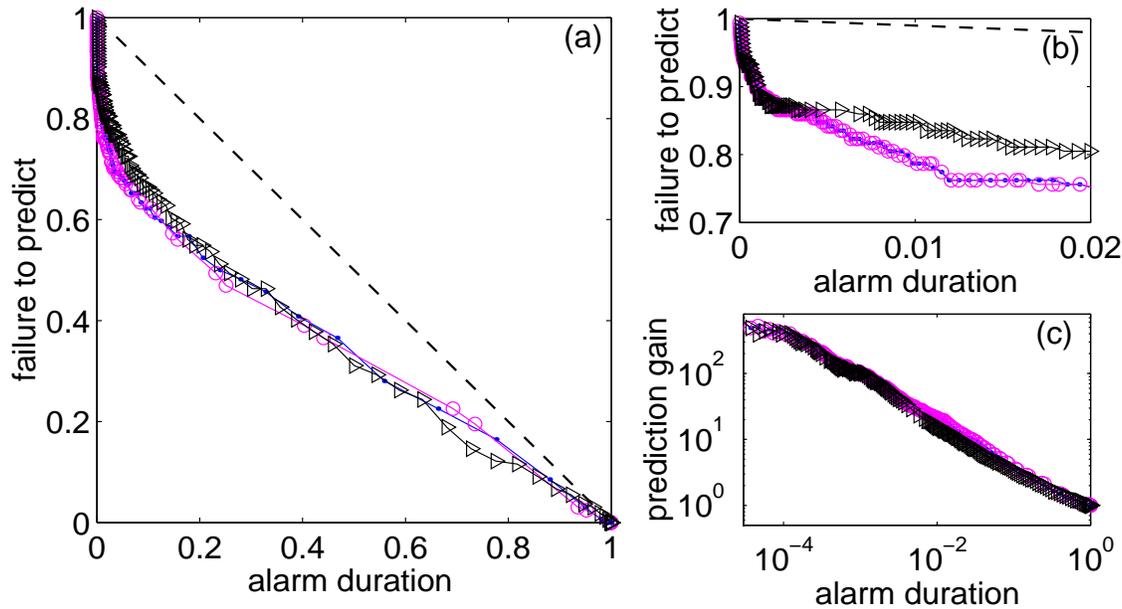,width=15cm} 
\caption{\label{erdiag3005T5}
Error diagram of different prediction algorithms, using either the
bare propagator (dots),
the median (circles) or the mean (triangles) number of events
obtained for the scenarios.
       The synthetic catalog and the prediction methods are the same as
for Figures \ref{Nt3005T5} and \ref{Nobspred3005T5}.
We use a time horizon (window size) of $T=5$ days to estimate the
seismicity rate
but we update the predictions each 0.5 day.
Target events are  $M\geq6$ earthquakes.
An alarm is defined when the predicted seismicity rate is above a threshold.
Each point of the curve corresponds to a different threshold ranging
from 0.1 to 1000
events per day.
The quality of the predictions is measured by plotting the ratio of
failures to predict
as a function of the total durations of the alarms normalized by the
duration of the catalog.
The results for these three algorithms are considerably better than
those obtained for a random prediction, shown as a dashed line for reference.
This Figure shows that about 20\% of large peaks of seismic activity
can be predicted
with a very small alarm duration of about 1\%. Panel (b) is a magnification
of panel (a) close to the origin of the alarm duration showing the very
fast increase of the success fraction ($=1-$ failure to predict) as the alarm
duration increases from $0$. Panel (c) shows that the predicted gain,
defined as the ratio of the success fraction over the alarm duration, decays
as a function
of the alarm duration from a very high value obtained for very short
alarm durations
as an approximate inverse power law with exponent slightly larger than
$1/2$.
}
\end{figure}

\clearpage

\begin{figure}
\psfig{file=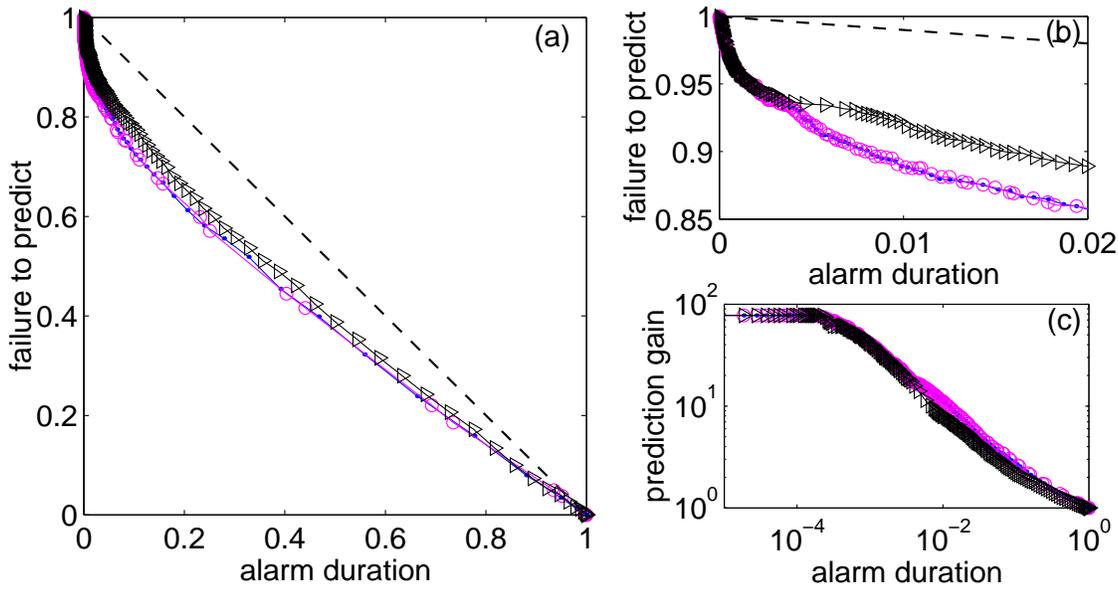,width=15cm} 
\caption{\label{erdiag3005T5M55} Same as Figure  \ref{erdiag3005T5}
but for targets with lower magnitudes $M \geq 5$.
Panel (c) shows that the predicted gain is again
approximately an inverse power law with exponent close to $1/2$ as a function
of the alarm duration.
}
\end{figure}


\begin{figure}
\psfig{file=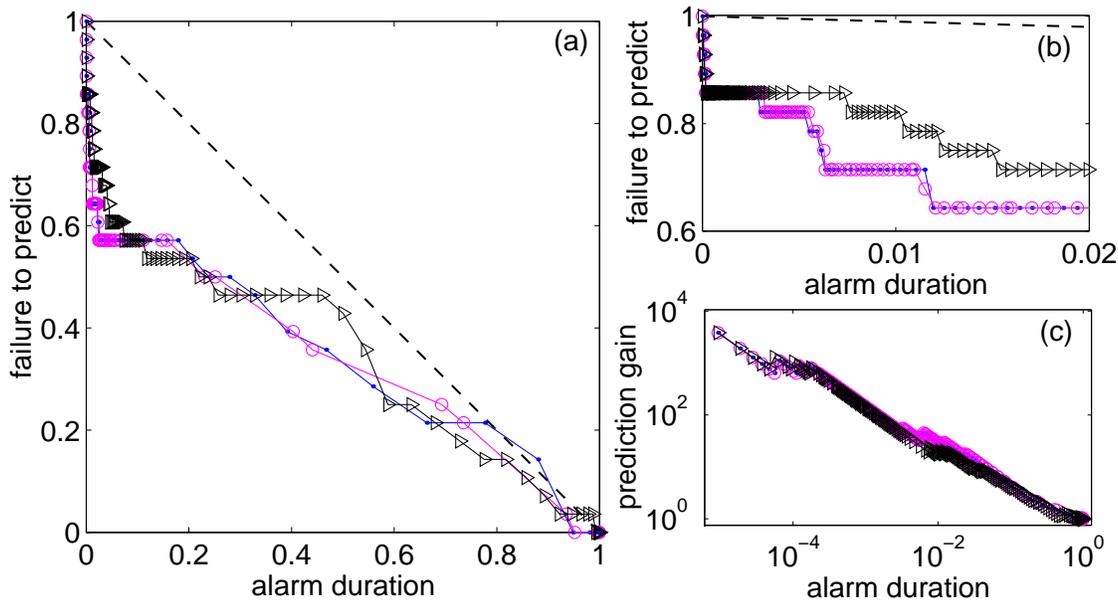,width=15cm}
\caption{\label{erdiag3005T5M7} Same as Figure  \ref{erdiag3005T5}
but for targets with larger magnitudes $M \geq 7$.
Panel (c) shows that the predicted gain is again
approximately an inverse power law with exponent slightly smaller
than $1$ as a function
of the alarm duration. Comparing this figure with Figures
\ref{erdiag3005T5} and
\ref{erdiag3005T5M55} suggests that the exponent defined in panel (c) is slowly
increasing with the magnitude of the targets.
}
\end{figure}

\clearpage

\begin{figure}
\psfig{file=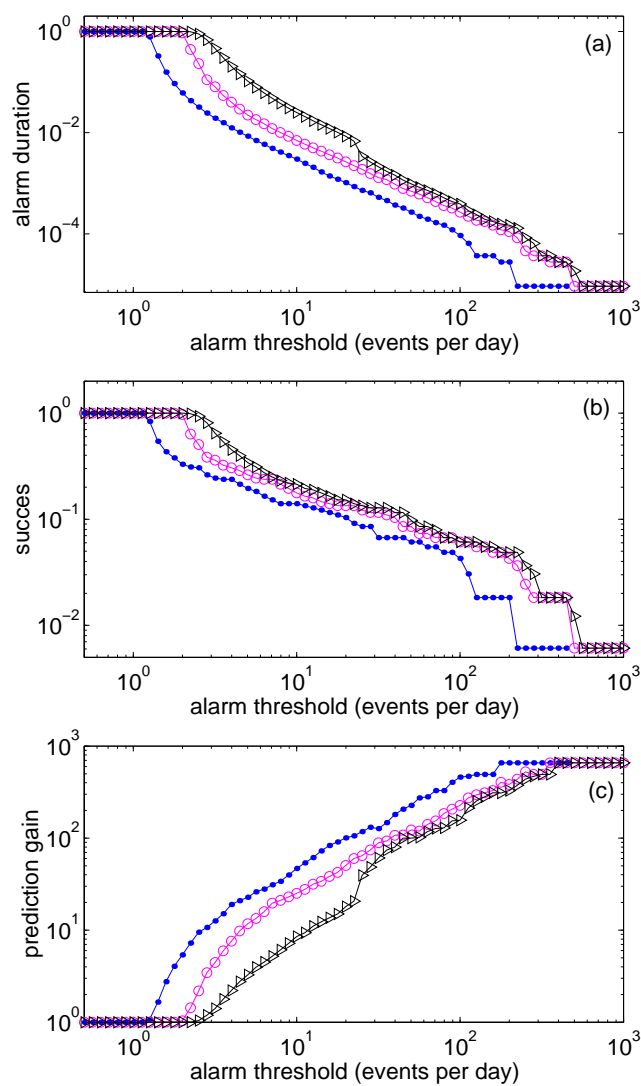,width=8.5cm} 
\caption{\label{dsg} Analysis of the prediction methods, using the
same synthetic catalog and predictions methods as for Figures
\ref{Nt3005T5}-\ref{erdiag3005T5M7}.
We use a time window $T=5$ days to estimate the predicted seismicity rate,
and a time $dT=0.5$ days between two updates of the prediction.
Target events are  $M\geq6$ earthquakes.
The duration of alarms normalized by the total duration of the
catalog is shown in panel (a) as a function of the alarm threshold
for the three
predictions methods : bare propagator (dots), the median (circles) and the mean
(triangles) number of events obtained for the scenarios.
The proportion of successes is shown in panel (b). The prediction
gain shown in panel (c) is defined by the ratio of the proportion of successes
(b) over the duration of alarms (a).
The prediction gain for large values of the alarm threshold is
significantly higher that the random prediction gain equal to $1$.}
\end{figure}

\clearpage

\begin{figure}
\psfig{file=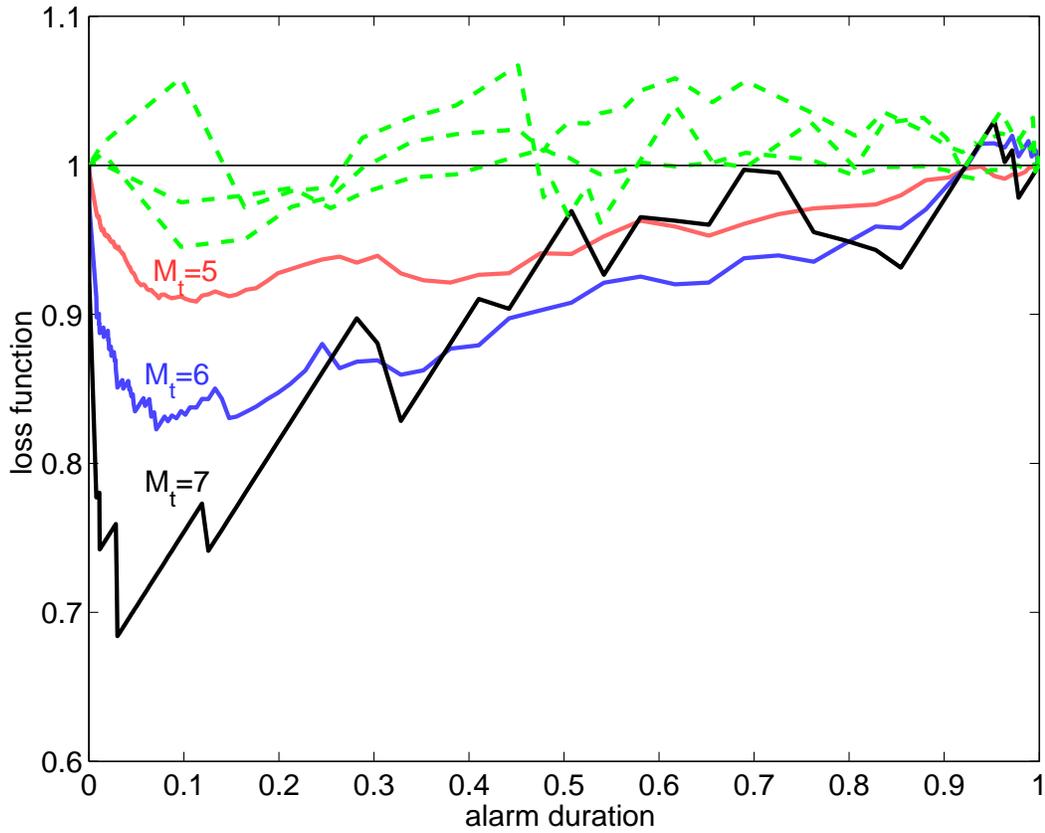,width=14cm} 
\caption{\label{dsgd} Loss function $\gamma$ defined as the sum
of the fraction of missed events and of the fraction of alarms.
We use the same catalog and prediction method as in Figures
\ref{Nt3005T5}-\ref{dsg}, with a time window of 5 days, and
an update time of 0.5 days. Solid lines give the results for
the prediction algorithm based on the ETAS model
(average of the seismicity rate over 100 scenarios) for different
values of the target magnitudes. Dashed lines correspond to the
predictions made using a Poisson process, with a seismicity rate equal
to the average seismicity rate of the catalog.}
\end{figure}

\clearpage

\begin{figure}
\psfig{file=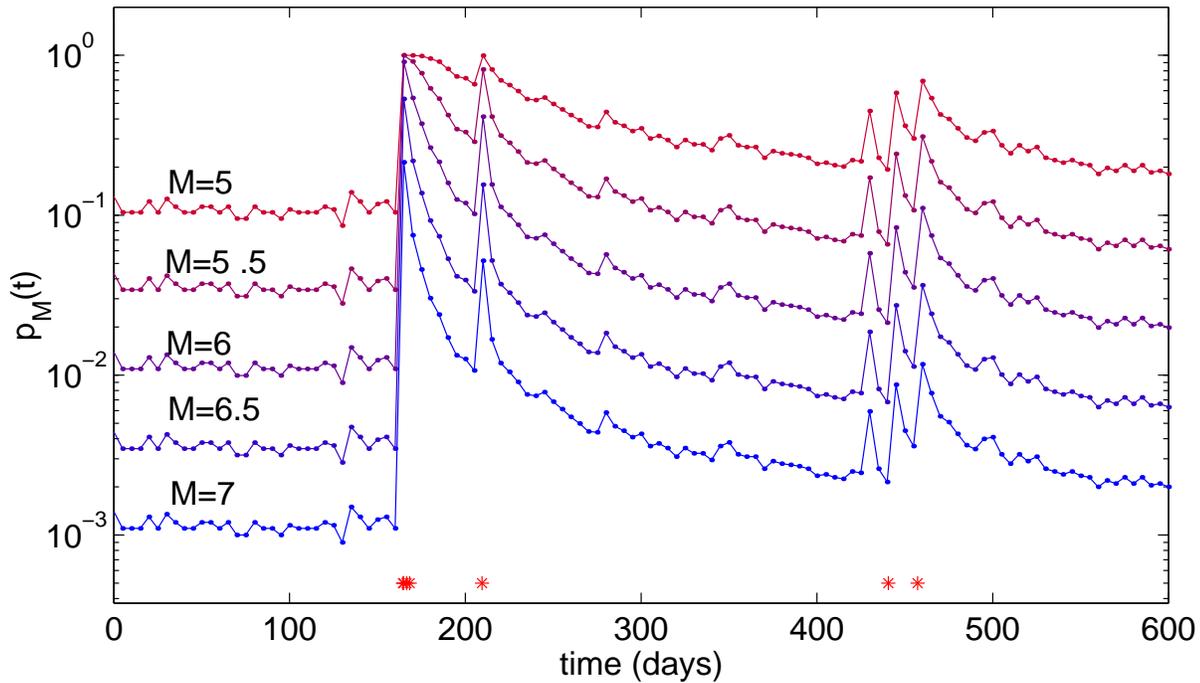,width=16cm}
\caption{\label{pimed}
Probability $p_M(t)$ of having at least an event of magnitude larger than $M$
in the time interval $(t,t+T)$ with T=5 days.
We have evaluated $p_M$ for $M$ between 5 and 7, from the seismicity rate
predicted by the median of 100 scenarios, and using equation (\ref{pi})
to estimate the probability $p_M(t)$ from the average seismicity rate
$\lambda(t)$ in the time interval $(t,t+T)$.
$p_M(t)$ is plotted for the sequence shown in Figure \ref{Nt3005T5}.
Stars indicate the occurrence of a large $M\geq 7$ earthquake.
The largest $M=8.55$ event of the sequence occurs at time $t=164$ days
when the seismicity rate is close to the average value. Thus, this event
cannot be predicted.
Six large $M>7$ earthquakes occur  in the following 500 days when the
seismicity
rate is still  above its average value, including three $M>7$ events
in the 5 days
immediately following the great event.}
\end{figure}

\clearpage

\begin{planotable}{ccrrrrrrrrrrrr}
\tablewidth{40pc}
\tablenum{1}
\tablecaption{\label{tabG} Prediction gain for different choices of
the alarm duration,
and/or different values of the time interval $T$, of the update time $dT$,
and of the target magnitude $M_t$.
$N_1$ is the number of targets $M\geq M_t$; $N_2$ is the number of intervals
with at least one target.
$G_{max}$ is the maximum prediction gain, which is realized for an alarm
duration $A$ (in proportion of the total duration of the catalog), which is
also given in the table.
All three prediction algorithms used here provide the same gain as a
function of the alarm duration,
corresponding to different choices of the alarm threshold on the
predicted seismicity rate.
$N_s$ is the number of successful predictions, using the alarm
threshold that provides
the maximum predictions gain $G_{max}$ for an alarm duration $A$ (we count only
one success when two events occur in the same interval).
This number $N_s$ is always very small, but a much larger number of successes
can be obtained with a larger alarm duration.
$N_{1\%}$, $N_{10\%}$, $N_{50\%}$ are the number of successes corresponding
to an alarm duration (in proportion of the total duration of the catalog)
of $1\%$, $10\%$ and $50\%$ respectively, corresponding to the prediction gains
$G_{1\%}$, $G_{10\%}$ and $G_{50\%}$. The values of $G_{50\%}$ show a
saturation
in the predictive power when increasing the fraction of alarm time,
reflecting the
fundamental limitation stemming from the fraction of large
earthquakes not associated
with a large seismic rate. Reading for instance of the last line of this table,
we observe that, out of 26 time windows of $50$ days that contained a $M\geq 7$
earthquake, we are able to predict 7 of them with only $1\%$ of the
time occupied
by alarms. Only two additional ones are predicted when using $10\%$ of the time
occupied by alarms. And only another four are predicted by increasing the time
of alarms to half the total duration of the catalog. The catalog spans
150 years corresponding to a little more than $10^5$ half-day periods.
}
\tablehead{\colhead{T } & \colhead{dT } &\colhead{$M_t$} & \colhead{$N_1$}
& \colhead{$N_2$} & \colhead{$G_{max}$} & \colhead{$A$} & \colhead{$N_s$}
      & \colhead{$N_{1\%}$} & \colhead{$N_{10\%}$}  & \colhead{$N_{50\%}$}
& \colhead{$G_{1\%}$}  & \colhead{$G_{10\%}$} & \colhead{$G_{50\%}$} \\
\colhead{ (days)} & \colhead{(days)}  }
\startdata
1.0 &1.0 &5.0 & 2003  &1332 &  40.4 & $3.2\times 10^{-4}$&
17&120&332&806&9.01&2.49&1.21 \nl
1.0 &1.0 &5.5 &  637  & 461 & 117.  & $7.4\times 10^{-5}$& 4 &
58&136&303&12.6&2.95&1.31\nl
1.0 &1.0 &6.0 &  198  & 159 & 339.  & $3.7\times 10^{-5}$& 2 & 30&
56& 94&18.9&3.52&1.18\nl
1.0 &1.0 &6.5 &   66  &  55 & 979.  & $1.9\times 10^{-5}$& 1 & 10&
15& 28&18.2&2.73&1.02\nl 
1.0 &1.0 &7.0 &   29  &  27 & 665.  & $5.6\times 10^{-5}$& 1 &  7&
11& 14&25.9&4.07&1.04\nl

5.0 & 0.5& 5.0 & 2003 &1389 & 77.5  & $1.1\times 10^{-4}$&12& 155&
382& 853& 11.2 &2.75& 1.23\nl
5.0 & 0.5& 5.5 &  637 & 483 & 223.  & $7.4\times 10^{-5}$&8 &  72&
155& 320& 14.9 &3.21& 1.33\nl
5.0 & 0.5& 6.0 &  198 & 164 & 656.  & $1.9\times 10^{-5}$&2 &  35&
64& 106& 21.3 &3.90& 1.29\nl
5.0 & 0.5& 6.5 &   66 &  57 & 1889. & $9.3\times 10^{-6}$&1 &  12&
18&  32& 21.0 &3.16& 1.12\nl
5.0 & 0.5& 7.0 &   29 &  28 & 3847. & $9.3\times 10^{-6}$&1 &   8&
12&  17& 28.6 &4.29& 1.21\nl

5.0 & 5.0& 5.0 & 2003 &1172 &  9.2  & $6.5\times 10^{-4}$&7 &
53&222&652&4.52&1.89&1.11\nl
5.0 & 5.0& 5.5 &  637 & 420 & 25.6  & $3.7\times 10^{-4}$&4 & 30&
93&253&7.14&2.21&1.20\nl
5.0 & 5.0& 6.0 &  198 & 145 & 74.3  & $2.8\times 10^{-4}$&3 & 16& 38&
85&11.0&2.62&1.17\nl
5.0 & 5.0& 6.5 &   66 &  53 & 203.  & $1.9\times 10^{-4}$&2 & 7 & 12&
30&13.2&2.26&1.13\nl
5.0 & 5.0& 7.0 &   29 &  26 & 414.  & $1.9\times 10^{-4}$&1 & 6 &  9&
14&23.1&3.46&1.08\nl

10.& 10.& 5.0 & 2003 &1067 &  5.1 & $5.6\times 10^{-4}$& 3
&32&167&584&3.00&1.57&1.09\nl
10.& 10.& 5.5 &  637 & 400 & 13.5 & $3.7\times 10^{-4}$& 2 &19&77
&229&4.75&1.93&1.15\nl
10.& 10.& 6.0 &  198 & 137 & 39.3 & $1.9\times 10^{-4}$& 1 &10&30 &
78&7.30&2.19&1.14\nl
10.& 10.& 6.5 &   66 &  50 & 107. & $1.9\times 10^{-4}$& 1 &5 &  8&
26&10.0&1.60&1.04\nl
10.& 10.& 7.0 &   29 &  24 & 224. & $1.9\times 10^{-4}$& 1 &5 &  7&
13&20.8&2.92&1.08\nl

50.& 50.& 5.0 & 2003 & 701 &  1.5 & 0.016  	        & 17
&11&84&370&1.57&1.20&1.06\nl
50.& 50.& 5.5 &  637 & 329 &  3.3 & $9.3\times 10^{-4}$& 1  &8
&43&181&2.43&1.31&1.10\nl
50.& 50.& 6.0 &  198 & 123 &  8.8 & $9.3\times 10^{-4}$& 1  &5 &20&
62&4.07&1.63&1.01\nl
50.& 50.& 6.5 &   66 &  48 & 22.4 & $9.3\times 10^{-4}$& 1  &4 &7 &
32&8.33&1.46&1.33\nl
50.& 50.& 7.0 &   29 &  22 & 48.9 & $9.3\times 10^{-4}$& 1  &4 &5 &
16&18.2&2.27&1.45\nl

50. &  5.& 5.0 & 2003 & 1172 & 9.2  & $6.5\times 10^{-4}$& 7
&53&209&657&3.37&1.78&1.12\nl
50. & 5. & 5.5 &  637 & 420 & 25.6  & $3.7\times 10^{-4}$& 4 &27&89
&251&4.76&2.12&1.20\nl
50. & 5. & 6.0 &  198 & 145 & 74.3  & $2.8\times 10^{-4}$& 3 &13&37 &
82&7.24&2.55&1.13\nl
50. & 5. & 6.5 &   66 &  53 & 203.  & $1.9\times 10^{-4}$& 2 &7 &11 &
24&8.49&2.08&0.91\nl
50. & 5. & 7.0 &   29 &  26 & 414.  & $1.9\times 10^{-4}$& 2 &7 &9  &
13&15.4&3.46&1.00\nl
\end{planotable}


\begin{planotable}{rrrrrr}
\tablewidth{22pc}
\tablenum{2}
\tablecaption{\label{tabgamma}
Minimum value of the error function $\gamma$ for each set of parameters 
$n$, $m_0$, $\alpha$ and $p$. Another measure of the predictability is 
the prediction gain ($G_{1\%}$) corresponding to an alarm duration of
$1\%$, which is equal to the percentage of predicted events. 
Predictions are done for the next day and updated each day ($T=dT=1$ days). }
\tablehead{\colhead{$n$} & \colhead{$m_0$} & \colhead{$\alpha$}
& \colhead{$p$} & \colhead{$\gamma$} & \colhead{$G_{1\%}$}}
\startdata
0.8  & 3.0 &  0.5 &  1.2 &  0.86  &  3.4\nl
0.8  & 3.0 &  0.8 &  1.2 &  0.84  &  8.9\nl
0.8  & 3.0 &  0.9 &  1.2 &  0.81  & 12.5\nl
0.5  & 3.0 &  0.8 &  1.2 &  0.89  &  4.8\nl
1.0  & 3.0 &  0.8 &  1.2 &  0.64  & 18.8\nl
0.8  & 1.0 &  0.8 &  1.2 &  0.83  &  6.5\nl
0.8  & 2.0 &  0.8 &  1.2 &  0.82  &  9.9\nl
0.8  & 3.0 &  0.8 &  1.1 &  0.86  &  9.3\nl
0.8  & 3.0 &  0.8 &  1.3 &  0.76  & 19.1\nl
\end{planotable}

\begin{planotable}{rrrrrrrrrrrr}
\tablewidth{40pc}
\tablenum{3}
\tablecaption{\label{tabB} Binomial scores $B$ for several prediction
algorithms and different choices of the time interval $T$ and the target 
magnitude $M_t$. We use non-overlapping time intervals for the
predictions  of length $T$ with a time
$dT = T$ between two predictions. $B_{med}$ is evaluated from the
median 
of the seismicity 
rate of the scenarios; $B_{mean}$ from the  average  seismicity rate; 
$B_{meanl}$  using the exponential $\lambda =\exp(\log <n>)$ of the 
average of the logarithm of the seismicity rate;
$B_{\phi}$ is measured using the bare propagator to estimate the
seismicity rate; $B_{pois_1}$ using a Poisson process with a seismicity 
rate equal to the average value of the catalog;
$B_{pois_2}$ using a Poisson process with a probability defined
as the fraction of intervals in the realized catalog that have
at least a target event. $N_1$ is the number of target events $M\geq M_t$ ; 
$N_2$ is the number of intervals with at least one target event. Note 
that $B_{med}$ seems to be often the best for the smaller magnitudes 
while $B_{mean}$ is often the best for the largest magnitudes. }
\tablehead{\colhead{T (days)} & \colhead{$M_t$} & \colhead{$N_1$}
& \colhead{$N_2$} & \colhead{$B_{med}$} & \colhead{$B_{meanl}$}
       &\colhead{ $B_{mean}$}  &\colhead{ $B_{\phi}$} &
\colhead{$B_{pois_1}$} & \colhead{$B_{pois_2}$}  }
\startdata
1. & 5.0 & 2003 &1332 &-5997.7& -5995.1& -6341.0& -6057.3& -6361.8 &-6243.4\nl
1. & 5.5 &  637 & 461 &-2512.4& -2511.7& -2614.6& -2545.0& -2678.7 &-2653.7\nl
1. & 6.0 &  198 & 159 &-1007.8& -1006.9& -1042.2& -1023.2& -1089.4 &-1085.0\nl
1. & 6.5 &   66 &  55 & -409.9&  -409.3&  -420.5&  -416.8&  -434.3 &-433.7\nl
1. & 7.0 &   29 &  27 & -217.8&  -216.6&  -218.1&  -224.0&  -233.3 &-232.1\nl
5. &5.0 & 2003 &1172& -3626.3& -3632.5& -3851.7& -3717.8& -3862.8 &-3705.5\nl
5. &5.5 &  637 & 420& -1720.0& -1719.1& -1774.8& -1765.4& -1810.7 &-1774.3\nl
5. &6.0 &  198 & 145&  -732.1&  -731.9&  -748.9&  -752.8&  -776.7 &-768.7\nl
5. &6.5 &   66 &  53&  -321.2&  -320.5&  -322.0&  -331.3&  -335.4 &-334.5\nl
5. &7.0 &   29 &  26&  -171.3&  -170.5&  -168.1& - 179.3&  -183.5 &-182.7\nl
10. &5.0 & 2003 &1067& -2651.0& -2662.7& -2822.4& -2736.0& -2852.4 & -2680.7\nl
10. &5.5 &  637 & 400& -1391.3& -1393.0& -1438.9& -1439.9& -1465.4 & -1424.7\nl
10. &6.0 &  198 & 137&  -621.1&  -621.2&  -637.3&  -640.3&  -648.6 & -638.2\nl
10. &6.5 &   66 &  50&  -276.6&  -276.2&  -280.1&  -286.1&  -285.2 & -283.7\nl
10. &7.0 &   29 &  24&  -147.2&  -146.4&  -145.3&  -155.3&  -154.3 & -153.9\nl
50. &5.0 & 2003 & 701&  -699.0&  -717.3&  -787.3&  -758.7&  -817.2 & -696.7\nl
50. &5.5 &  637 & 329&  -658.9&  -666.0&  -698.8&  -702.7&  -706.2 & -662.8\nl
50. &6.0 &  198 & 123&  -379.6&  -381.1&  -392.5&  -398.5&  -395.5 & -382.6\nl
50. &6.5 &   66 &  48&  -192.2&  -191.5&  -192.5&  -204.3&  -197.9 & -196.2\nl
50. &7.0 &   29 &  22&  -104.5&  -103.5&  -102.3&  -113.3&  -107.5 & -107.4\nl
\end{planotable}

\end{document}